# Nanoporous ionic organic networks: from synthesis to materials applications

Jian-Ke Sun, Markus Antonietti and Jiayin Yuan*

Max Planck Institute of Colloids and Interfaces, Department of Colloid Chemistry, D-14424 Potsdam, Germany, jiayin.yuan@mpikg.mpg.de

**Abstract:** The past decade has witnessed the rapid progress in synthesizing nanoporous organic networks or polymer frameworks for various potential applications. Generally speaking, functionalization of porous networks to add extra properties and enhance materials performance could be achieved either during the pore formation (thus a concurrent approach) or post-synthetic modification (a sequential approach). Nanoporous organic networks which include ion pairs in a covalent manner are of special importance and possess extreme application profiles. Within these nanoporous ionic organic networks (NIONs), here with a pore size in the range from sub-1 nm to 100 nm, we observe a synergistic coupling of the electrostatic interaction of charges, the nanoconfinement within pores and the addressable functional units in soft matter resulting in a wide variety of functions and applications, above all catalysis, energy storage and conversion, as well as environmental operations. This review aims to highlight the recent progress in this area, and seeks to raise original perspectives that will stimulate future advancements at both the fundamental and applied level.

## 1. Introduction

### 1.1 General

Currently, there is a general consensus that novel materials are needed to address renewable energy storage and conversion technologies and environmental remediation processes for a more sustainable development of our planet earth.[1-3] To improve the efficiency and meet the ever-increasing energy requirements in various systems, nanoporous materials have been regarded as one of potential candidates due to the intrinsic characteristic of open channels and large specific surface area, coupled to the possible control of the accessibility, percolation as well as optimized mass transport for various device applications.[4-11] Among the developed porous materials, polymer-based species are attractive, as the easy processing and wide variability on monomers enabled advanced engineering within rather simple approaches. Typical features of those systems are high specific surface area, diverse pore dimensions, the use of lightweight elements only, strong covalent linkages, as well as addressable chemical functions.[12-15]

Nanoporous materials refer to a class of porous materials having pore diameters from 100 nm down to around 1 nm. According to the international union of pure and applied chemistry (IUPAC) definition, micro-/meso-/macropores cover a pore size range of <2nm, 2-50 and >50 nm, respectively. Though "nanoporous materials" is not a strict term defined by the IUPAC nomenclature, it has been widely employed in the field of materials science and is used here accordingly. Traditional porous polymeric materials, derived from classical polymeric chemistry have a pore size mainly in the mesopore and macropore range, and are only partially considered in this review (<100 nm). These classical systems are fabricated commonly by the phase separation,[16] emulsion polymerization[17] or the hard template method.[18] On the other side, the recently developed microporous organic polymers (MOPs) include in fact several subclasses, such as polymer of intrinsic microporosity (PIMs),[15] covalent organic frameworks (COFs),[19] porous aromatic frameworks (PAFs),[20] and conjugated microporous polymers (CMPs).[14] They are conventionally synthesized by bottom-up approach from stiff molecular building blocks and usually possess a very high surface area as well as ordered pore architecture in some cases, *e.g.*, COFs.

Functionalization of such porous polymers is usually target-motivated, that is key properties are adjusted and optimized to serve a specific purpose. Broadly speaking, the functions could be classified as chemical and physical ones. Examples of the former are acidity/basicity,[21-22] the ability to coordinate, or chemical activity under certain conditions and stimuli.[23-26] Typical examples of the latter are the electrical or optical properties.[27-31] Indeed, by means of versatility of organic chemistry, various functional moieties can be incorporated by direct synthesis from functional monomers or by the post-synthetic modifications (PSM) of the prefabricated porous skeleton. It should be noted that, for some porous matrices, the incorporation of additional chemical groups onto the pore wall will block the pores or at least restrict pore accessibility and lower pore volume and specific surface area.[32-35] This is a general phenomenon, *e.g.* in silica, zeolites, or metal-organic frameworks (MOFs). In this context, we emphasize that the incorporation of charged species into the porous skeleton could lower steric problems, because the dynamic ionic bonds between host skeleton and counterions add a dynamic component to host-guest interactions and the blocking effect.[36-41]

## 1.2 Nanoporous ionic organic networks (NIONs): classifications and brief overview

NIONs are obtained when the polymer network contains extra charges of one type, with free counterions, either positive or negative, electrostatically bound to maintain the overall electrical neutrality. Compared with neutral polymer networks, NIONs carry extra possibilities to engineer and fine-tune advanced functional porous materials. In addition to dynamic blocking or gating, as well as the ability to readjust the pore size by counterion exchange and counterion mixing, the existence of charge on the pore wall endows the pore skeleton with selective interactions with guest molecules due to the intrinsic charge repulsion/affinity effects, *e.g.* by Coulomb interaction, but also by ion-π-interactions or ion bridges. Electrostatic interaction therefore can play a critical role in amplifying separation and sorption efficiency, and the decorated ions in the porous networks could be chosen to increase the adsorbate-adsorbent interaction through the polarization effects/chemical bonding.[37] An enhanced separation efficiency can arise when the pore size drops below 10 nm so that the nanoconfinement effects allow a stronger coupling with the electrostatic interaction as a function of pore size.[42-49] For macroporous scaffolds, Coulombic charge plays a crucial role in maintaining appropriate hydrophilicity of scaffolds and affecting cell adhesion and recognition. For nanofiltration membranes, the surface charge as well as in many cases chemical specificity is important in controlling salt-rejection performances. Last but not the least, the incorporation of charged units into conjugated porous skeletons can promote or bias the electron/hole mobility in the skeleton due to the localization effects,[50-52] which leads to intriguing photoelectrochemical effects for energy applications.

The NIONs reviewed here involve the positioning of charges within the scaffold through top-down, bottom-up and post-synthetic methods. Partially or weakly charged porous networks as well as neutral skeletons with free ionic pairs or salts are out of the scope of the present review. We meanwhile try to select cases which feature how charge on the scaffold affects the properties of materials, while possibly analyzing the relationship between the charged character and pore size.

## 2. Synthesis of NIONs

The past several years have witnessed the rapid development of various synthetic strategies towards NIONs, which include the hard/soft templating method, the direct synthesis of microporous ionic organic network, the free radical polymerization approach, the ionic complexation method as well as the post-synthetic modification. Each strategy demonstrates distinct advantages with respect to the pore structure control, processing of porous materials, or scalable production, as well as limitations from a synthetic perspective. A flexible balance of the porous structural parameters (pore shape, pore size, pore size distribution, pore distribution profile, *etc.*) is thus to be considered as a compromise to satisfy a specific research or application goal. Table 1 summarizes the strength and weakness of these strategies.

**Table 1**. Summary of current methods for the preparation of NIONs

| Methodology | Advantages | Drawbacks |
|---|---|---|
| Hard templating | Easy synthesis of porous networks with controlled meso/macropore size due to structurally well-defined porous template | Sacrificial templates are needed and later removed, scale-up production is difficult and limited utility of materials; the products lack micropores |
| Soft templating | Polymers bear tailored pore size usually in meso/macropore range; well-defined pore architecture can be obtained | Control of interactions between the ionic organic precursor and copolymer template is tricky; the products lack micropores |
| Direct synthesis of microporous ionic organic network | Permanent porosity with high surface area is achievable; the pore structure/functional ionic sites can be tailored by designing monomers with targeted structures; ordered crystalline ionic network is achievable by judicious choice of building blocks and the reaction types | Strict requirements on the monomer structures and synthetic routes are needed; fine design and control over meso- /macropores is unavailable; processing of materials for device applications is still a challenge |
| Free radical | Easy fabrication procedure without template, the | Difficult to delicately tailor the pore |
| polymerization | product contains a broad range of pore size from micropores to meso/macropores; large scale production is possible. | structure |
| Ionic complexation | Simple and large-scale synthetic method for meso/macroporous networks without template; pore size/structure can be controlled by tuning the species of polyelectrolytes; porous membrane synthesis is possible | Not stable in highly concentrated ionic solution; pore structure lacks micropores, moderate BET surface area |
| Post-synthesis | Abundant porous network precursors are available for post-modification; porous structure can be processed with custom-designed ionic functionalities | Homogeneous grafting of ionic moieties onto prefabricated skeleton is difficult in some cases; serious decrease of pore accessibility/structure distortion occurs if parent porous skeleton is instable in the post-synthesis condition |

3.

## 2.1 Templated synthesis of NIONs

The template method is conceptionally straightforward and has been the most extensively employed for the preparation of porous networks ranging from inorganic to organic materials or their hybrids. Ideally, it is a kind of molding or casting technique for the direct replication of the inverse structure of a prefabricated template with shaped morphology. *i.e.* the template defines the pore. For the preparation of NIONs, compatibility between the ionic precursor, the host matrix and the template needs to be carefully considered. Aiming at a faithful template leading to a high quality of the porous polymer network, surface modifications are necessary in some cases to accommodate porous active sites and guide the growth of ionic polymers. Besides, the template should be stable enough to sustain the conditions employed in the polymerization or polymer formation process. Last but not the least, a high degree of crosslinking is a "must" to trap the polymer structure against chain motion and pore collapse after template removal.

### 2.1.1 Hard-template synthesis of NIONs

#### 2.1.1.1 Infiltration method

The infiltration method is based on the use of porous inorganic materials or nanostructured materials as templates. Typically NIONs could be obtained *via in situ* polymerization of ionic monomers which are previously infiltrated inside the cavities of the porous inorganics, and the subsequent removal of the template. This method generates NIONs with meso- and macropores as a result of easy filtration of such pores by polar monomers. A particular family of ionic monomers is ionic liquids (ILs), substances consisting entirely of ions and being liquid below 100 $^{o}C$. Polycondensation or polymerisation of fluidic IL monomers into porous materials yields porous IL networks that integrate some unique characteristics of ILs into porous materials.[53-55] Readers interested in porous networks made up from crosslinked ILs are referred to a recent, elegant and brief review by Watanabe *et al.*.[11] The pore structures obtained by this method are dictated by the templates if the infiltration process is complete, which is sometimes assisted by vacuum treatment. For example, Li *et al.* developed a series of porous poly(ionic liquid)s (PILs) by using porous inorganic materials (silica or $CaCO_3$) as hard templates.[56,57] The procedure followed a three-step approach: first fabrication of a silica template, second infiltration of the template with a mixed solution of an IL monomer, a dimethacrylate crosslinker and an initiator followed by polymerization at an elevated temperature, and third selective dissolution of the silica hard template to afford the porous structures (inverse opal). Porous crosslinked PILs with ordered or disordered structures and the pore size ranging from meso- to macropore could be achieved through appropriate choice of the template precursor. The obtained ordered materials show intriguing applications depending on the size and ordering of the pores. The 3D ordered macroporous PILs with pore size ranging from tens of nanometers to several hundreds of nanometers could be used as optical sensors to recognize anions, sense humidity and determine solvent polarity by colour variation of the porous PIL film.[55-57] When reducing the pore size to below 50 nm by using smaller silica nanoparticles as hard template, the mesoporous PIL frameworks are useful for gas capture. In a recent example, a mesoporous PIL was synthesized by using densely packed silica nanoparticles (nominal diameter: 25 nm) as template. The IL monomer 3-(4-vinylbenzyl)-1-vinylimidazolium bis(trifluoromethane sulfonyl)imide was synthesized *via* quaternization reaction of 4-vinylbenzyl chloride with 1-vinylimidazole, followed by anion exchange to replace the chloride with large-sized fluorinated anion bis(trifluoromethane sulfonyl)imide ($Tf_2N$, also often abbreviated as TFSI in literature) (Fig. 1) to generate the room-temperature ionic liquid (RTIL) monomer.[58] The Brunauer–Emmett–Teller (BET) specific surface area ($S_{BET}$) and average pore size were found to be 220 $m^2$ $g^{-1}$ and 15 nm, respectively (note that though $S_{BET}$ is suitable for mesopores and not for micropores, it is nevertheless widely used to compare various materials.). The intrinsic affinity of the IL species to $CO_2$ and the stable mesopore transport structure made the material a fast and effective $CO_2$ sorbent**.**

It is worth mentioning that rational functionalization of pore walls facilitates infiltration of ionic species. One example is the copolymerization of an IL 3-benzyl-1-vinylimidazolium bromide with divinylbenzene as cross-linker in the presence of O-silylated SBA-15 (average pore size: 10.5 nm) as the hard template.[59] The surface hydroxyl groups in SBA-15 were protected by trimethylsilyl groups to assist diffusion of the hydrophobic IL monomer into the pores for nanocasting. The obtained material gave $S_{BET}$ and average pore size of 289 $m^2$ $g^{-1}$ and 2.4 nm, respectively. In addition, rational adjustment of chain conformation can improve macromolecule infiltration. For example, as a weak polyelectrolyte poly(acrylic acid) (PAA) chains at a low pH or high ionic strength exhibited a coiled conformation and can infiltrate the nanoporous silica template, while they adopted an extended chain conformation at a high pH or low ionic strength, thus spatially excluded from the nanopores.[6] Currently the infiltration method has been only applied with silica and $CaCO_3$ template, but its principle can be extended to other inorganic ones, such as titania or alumnia.

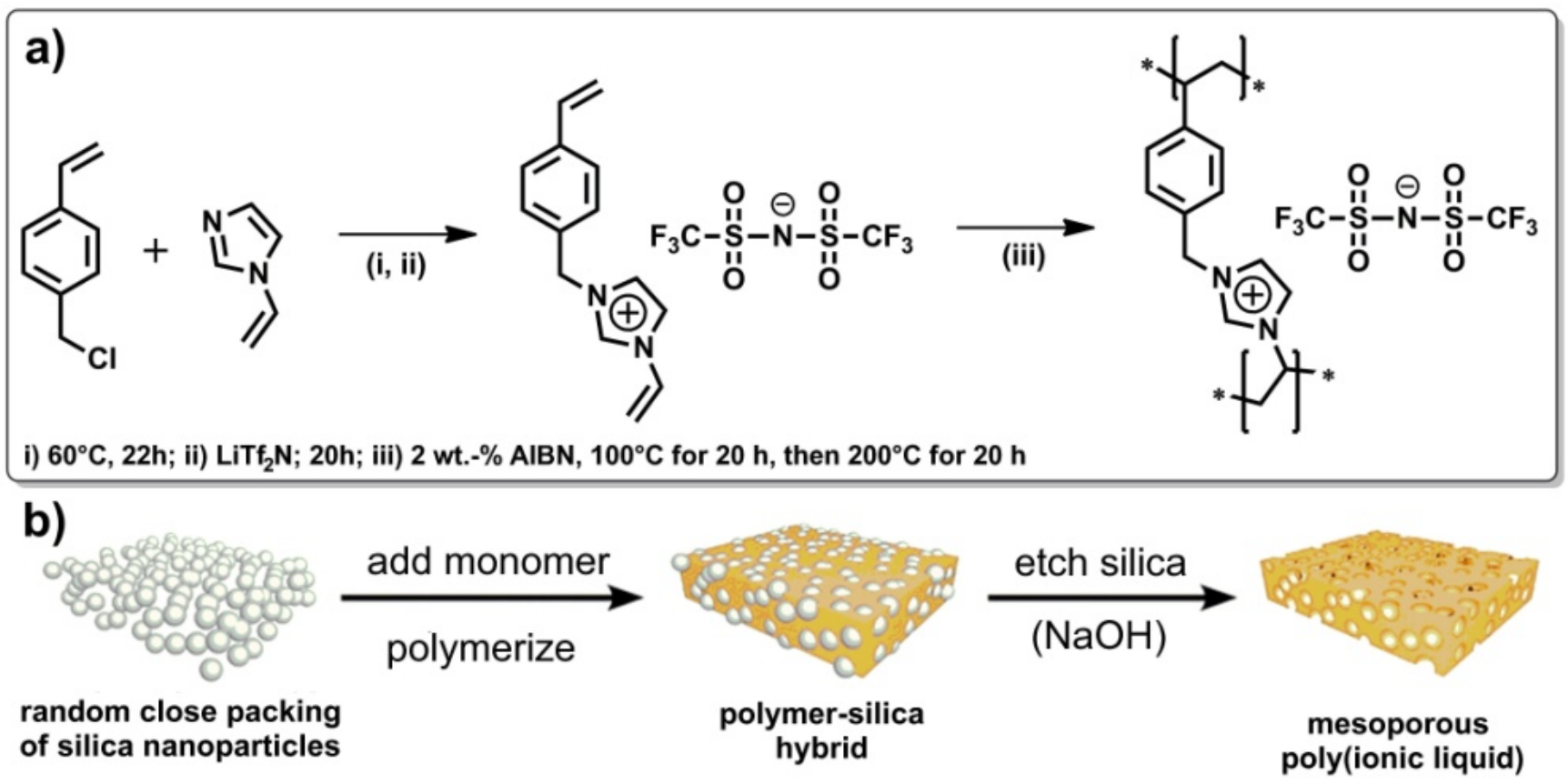

**Fig. 1** (a) Synthetic pathway towards the mesoporous PIL and (b) schematic overview of the employed hard-templating pathway. Reprinted with permission from ref. 58. Copyright 2012 American Chemical Society.

### 2.1.1.2 Layer-by-layer (LBL) assembly

Layer-by-layer (LBL) technique is able to engineer ionic polymers into NIONs, i.e. a *top-down* method. The LBL method is conducted by depositing the charged polymers onto the surface of porous templates *via* alternating adsorption of oppositely charged polyelectrolytes.[60-62] In this regard, the polycation and polyanion could complex under electrostatic attraction, and the subsequent removal of the template gives rise to porous polymeric materials crosslinked by Coulomb forces.[63] This technique peculiarly allows for precise control of the thickness of the polymer layer as well as the surface morphology of the template in the nanometer range. The LBL technique is applicable to a variety of templates, *e.g.* inorganic solid spheres, anodic aluminium oxide (AAO), mesoporous silicas, porous fibers, rods, *etc.* and features a large structure toolbox due to combination of various polycations and polyanions.

Although the principle to construct ionic porous network by LBL technique is conceptually simple and straightforward, attention needs to be paid to processing conditions, including ionic strength, solution pH and polymer concentration since these factors control the physicochemical processes at and in the templates. In particular, charge regulation within nanopores as large as 10 times of the diameter of the deposition species can lead to inhibited pore partitioning (*i.e.* pore clogging).[64]

For LBL-derived NIONs, also pore stability is a challenge since the ionic interactions stabilizing the porous network are easily affected by external conditions. For example, the electrostatic force is inversely proportional to the dielectric constant of the environmental solvent;[65] weak polyelectrolyte pairs based NIONs with weaker ionic interaction are prone to reconformation, and the presence of highly concentrated salt solutions leads to possible disassociation of the material. To strengthen the pore stability, other methods have been co-employed such as interlayer crosslinking of the ionic polymer with additional crosslinkers, covalent bonds or post-treatment.[6] Caruso *et al.* reported a selective etching technique to prepare nanoporous ionic films.[66] In their work, the films comprising three different components, poly(allylamine hydrochloride) (PAH) as polycation, PAA as weak polyanion, and poly(4-vinylpyridine) (P4VP) as hydrogen-bonding polymer (also a weak polycation) were assembled on nonporous planar templates at pH 3.5, after which two of the components (PAH and PAA) were stabilized *via* covalent crosslinking through formation of amide linkage at elevated temperature. The pH was then elevated to 10 to disrupt the hydrogen-bonding between P4VP and PAA, and hence removed the sacrificial component (P4VP), resulting in nanopores (diameter 10–50 nm) within the multilayer film (Fig. 2).

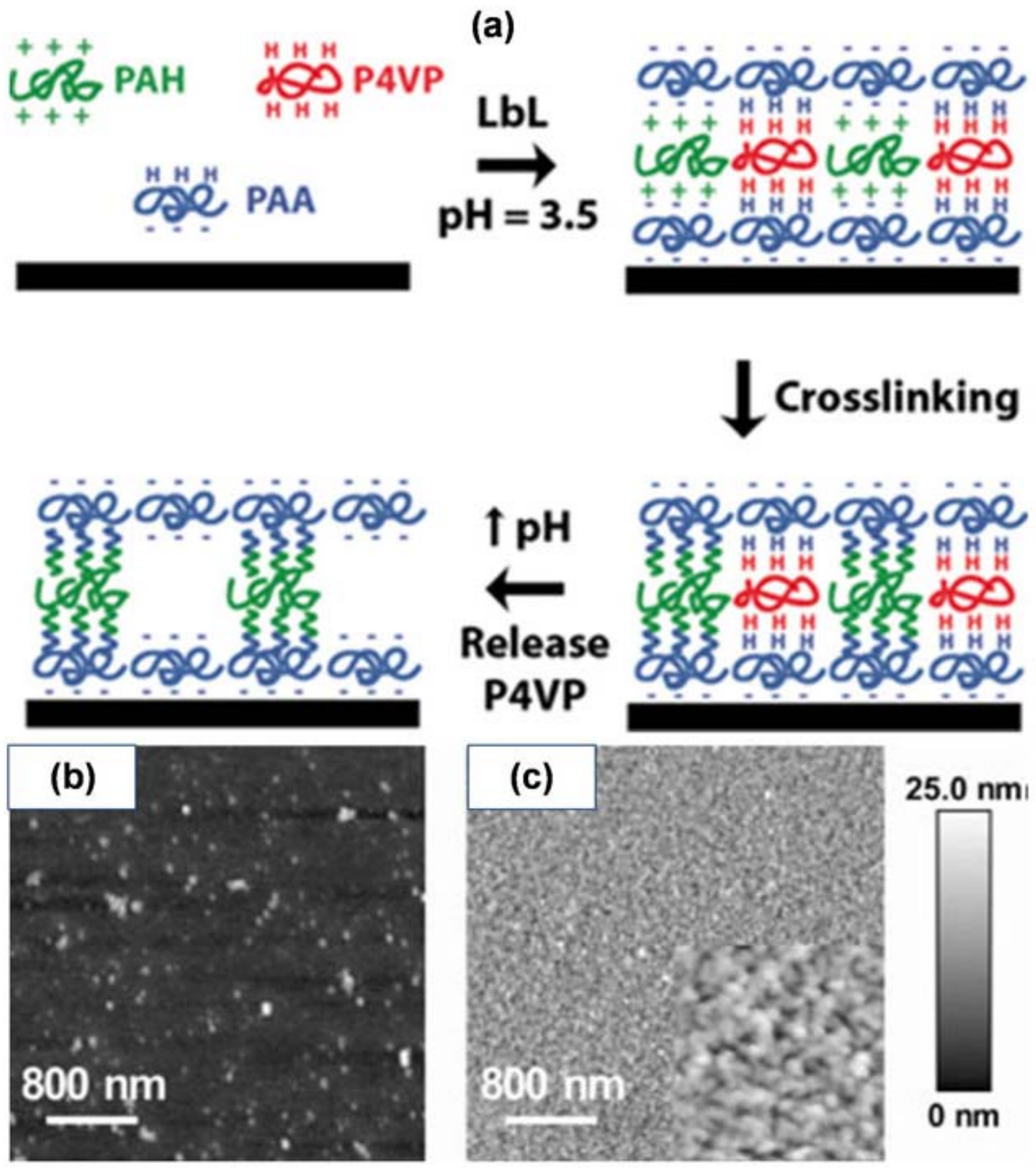

**Fig. 2** (a) Procedure used to produce porous polymer films from PAA/(P4VP/PAH) multilayers. Initially, a multilayer thin film is formed by adsorbing PAA in alternation with a blend of P4VP and PAH. The film is then thermally crosslinked *via* formation of covalent amide linkages between PAA and PAH molecules. Finally, P4VP is released by elevating the pH to disrupt hydrogen bonding between PAA and P4VP. AFM images of the $[PAA/P4VP(75)/PAH(25)]_{10}$ multilayer films prepared at pH 3.5 before (b) and after (c) crosslinking followed by exposure to a pH 10 buffer solution. The inset displays a higher magnification (500 nm × 500 nm) of the pores induced in the multilayer film. Reprinted with permission from ref. 66. Copyright 2005 Wiley-VCH.

Very recently, Ogoshi *et al.* developed a unique strategy to introduce even smaller pores through construction of pillar[5]arene-based nanoporous ionic film (Fig. 3).[67] Their work employed LBL assembly with consecutive adsorption of cationic and anionic pillar[5]arenes. The resultant pore size was around 5 Å, which is the inherent cavity of pillar[5]arene molecules. The films featuring active pores allowed for shape-selective uptake of dinitrobenzene isomers: the film adsorbed para-dinitrobenzene but rejected ortho- and meta-dinitrobenzene. This selectivity was also bound to the surface electrostatic potential: para-dinitrobenzene was adsorbed into the films with a positive surface, but not the negative one.

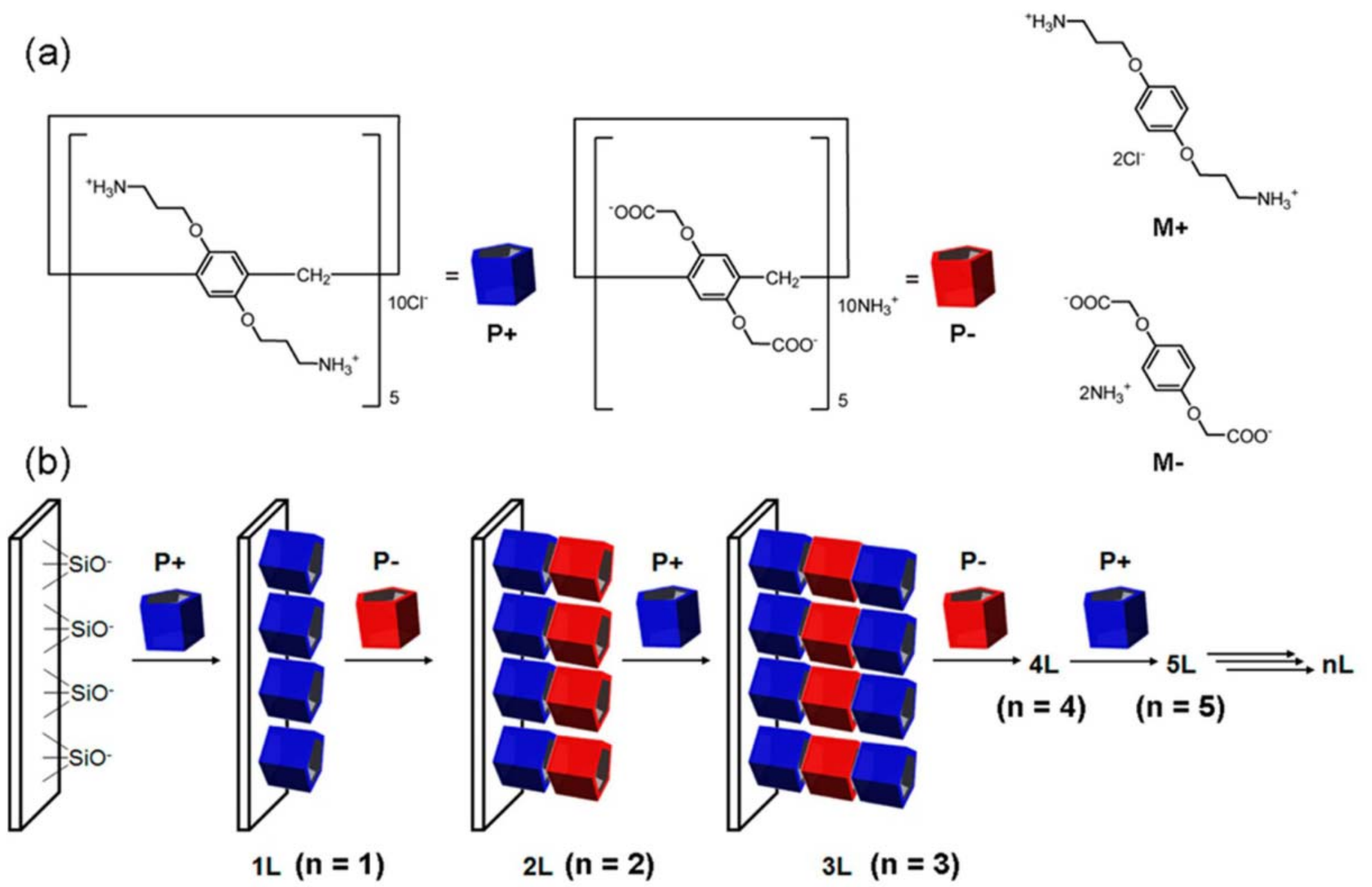

**Fig. 3** (a) Chemical structures of cationic (P+) and anionic (P−) pillar[5]arenes and cationic (M+) and anionic unit (M−) models. (b) LBL assembly by consecutive adsorption of P+ and P−. Reprinted with permission from ref. 67. Copyright 2015 American Chemical Society.

LBL technique may combine other processing methods to introduce pores of different length scale beyond the template. An example of preparation of hollow ionic polymeric capsules (PAH/PSS) (PSS = poly(sodium 4-styrene sulfonate)) by combination of the LBL technique and a freeze-drying process was reported by Schönhoff *et al.*[68] The holes with an average size of 16 nm could be generated owing to the osmotic pressure difference in the core dissolution process.

### 2.1.2 Soft template synthesis of NIONs

The soft template method is equally important in the synthesis of porous materials. The method is based on the use of soft matters, majorly self-assembled block copolymers, as template. The phase separation occurs in the self-assembly process since the thermodynamic incompatibility of compositional heterogeneous segments prefers to minimize contact energy. Because of segment chain connectivity, the separation is restricted to the nanometer scale to form ordered structural patterns. Generally, the resultant materials are dominantly meso- and macroporous, that is, the pore size is well above that of a single polymer chain. Especially materials with well-defined ordered mesopore structures were obtained on the basis of this method. Although soft templating has been well explored for the synthesis of porous inorganic materials or neutral organic materials or their hybrids,[69-71] the exploration of this method towards ionic polymer frameworks is seldomly reported. Matching the electrostatic charge-density is important for enthalpic reasons. The interaction between the ionic organic precursor and copolymer template on one side must be strong enough to spatially guide the polymerization around the template, and on the other side should be weak enough that the template could be fully removed to reopen the porous structure. Meanwhile, entropic interactions as a key player here restrict polymer chain conformations in a spatially confined reaction field.

Wang and co-authors reported an interesting example to synthesize NIONs by balancing the interaction between the soft template and the ionic polymers. In their work, a hierarchical meso-/macroporous PIL monolith with tuneable pore structure was synthesized through free radical polymerization of an IL (1-allyl-3-vinylimidazolium chloride) by using the triblock copolymer P123 ($EO_{20}PO_{70}EO_{20}$) as the soft structure-directing template.[72] The dissolved P123 and the IL effectively interacted with each other through the $S^0H^+X^-I^+$ mode ($S^0$: nonionic surfactant P123, $H^+$: hydrogen ions of ionization, X: IL anion, $I^+$: IL cation), in which the $H^+$ concentration increases after introduction of the initiator ammonium persulfate. Protons were necessary to realize the interaction between the ionic polymer and the soft template P123, while the template can be removed by solvent extraction of the as-synthesized material in ethanol to leave behind the pores. The product possessed a $S_{BET}$ and average pore size of 143 $m^2$ $g^{-1}$ and 28 nm, respectively. Similarly, Xiao and co-workers synthesized mesoporous sulfonated melamine–formaldehyde resin assisted by copolymer surfactant F127 as template.[73] After removal of the copolymer surfactant by ethanol extraction, the NION exhibited uniform mesopores with a $S_{BET}$ and average pore size of 256 $m^2$ $g^{-1}$ and 10.2 nm, respectively.

## 2.2 Template-free synthesis of NIONs

From a synthetic point of view the template method is popular and straightforward, the process nevertheless inevitably involves the synthesis and removal of sacrificial components, which is time-/energy-consuming and in some cases comes with non-sustainable steps. Besides, the template method is powerful to produce meso- and macroporous NIONs, but significantly restricted when microporous materials are concerned. The introduction of both porosity and charge into organic "soft" materials without templates imposes strict demands on the resulting framework, as polymer chains normally pack space-efficiently to maximize intermolecular interactions, especially when electrostatic interactions between ionic species in the skeletons are involved. As a rule of thumb, the resulting polymers should be both very stiff and badly packing to enable such spontaneous porosity. Fortunately, diverse available functional groups and refined covalent-bond forming reactions in organic synthesis provide opportunities to tailor the pore structures of NIONs under such template-free conditions. Table 2 summarizes all the NIONs discussed in the following section prepared *via* the template-free approach.

**Table 2** Texture parameters of typical NIONs from template-free method

| NION | $S_{BET}$ ($m^2/g$) | Pore size (nm) | Pore volume ($cm^3/g$) | Synthesis method | Ref. |
|---|---|---|---|---|---|
| Ru-PCP | 1348 | 0.75-1.65 | – | Alkyne trimerization | 24 |
| Ir-PCP | 1547 | 0.75-1.65 | – | Alkyne trimerization | 24 |
| Li-ABN | 890 | – | 0.61 | Sonogashira coupling | 40 |
| ICOF-1 | 1022 | 1.1 | – | Spiroborate–linked condensation | 50 |
| ICOF-2 | 1258 | 2.2 | – | Spiroborate–linked condensation | 50 |
| Tetraarylborate polymer network-16 | 480 | – | – | Sonogashira coupling | 82 |
| PAF-23 | 82 | 0.44 | – | Sonogashira coupling | 83 |
| PAF-24 | 136 | 0.50 | – | Sonogashira coupling | 83 |
| PAF-25 | 262 | 0.47 | – | Sonogashira coupling | 83 |
| PP-Br | 650 | ~ 0.7 and 1.4 nm | – | Yamamoto-type cross-coupling | 84 |
| Hypercrosslinked phosphonium polymer-1 | 1168 | 1.18 | 1.00 | Friedel–Crafts reactions | 85 |
| CCLIL-beti | 814 | < 2 | 0.36 | Polycondensation | 86 |
| CCLIL-$Tf_2N$ | 481 | < 2 | 0.08 | Polycondensation | 86 |
| T-IM | 620 | < 2 | 0.36 | Sonogashira coupling | 87 |
| EOF-16 | 262 | < 2 | 0.21 | Suzuki coupling | 88 |
| PCP-Cl | 755 | < 2 | – | Sonogashira coupling | 89 |
| PCP-$BF_4$ | 586 | < 2 | – | Sonogashira coupling | 89 |
| PCP -$PF_6$ | 433 | < 2 | – | Sonogashira coupling | 89 |
| POP-V1 | 812 | 0.6 | – | Sonogashira coupling | 90 |
| POP-V2 | 960 | 1.25 | – | Sonogashira coupling | 90 |
| EB-COF:Br | 616 | 1.66 | – | Schiff base reaction | 91 |
| PCu–NHC | 62 | < 2 | 0.18 | Carbene-metal complexation | 92 |
| Au-NHC@POPs1 | 798 | ~ 1 | – | Sonogashira coupling | 93 |
| Poly-NHC-2–$Pd^{2+}$ | 569 | < 2 | 0.09 | Carbene-metal complexation | 94 |
| BPP-7 | 705 | ~0.6 | – | Suzuki coupling | 170 |
| Polymer network 7 | 386 | < 2 | – | Oxidative coupling | 97 |
| PCIF-1 | 1025 | 3.52 | 0.9 | Nucleophilic substitution reaction | 99 |
| PAF-50 | 384 | 0.5 | – | Nucleophilic substitution and condensation reaction | 100 |
| PIL1:1 | 86 | 6.8-8.5 | 0.22 | Free radical homopolymerization | 102 |
| MesoPIL-1 | 429 | 3.6 | 0.38 | Free radical homopolymerization | 104 |
| P(DVB-0.1BVI$PF_6$) | 935 | 10-50 | 0.77 | Free radical copolymerization | 106 |
| HCP-IL-2 | 647 | 2.5 | 0.28 | Free radical copolymerization and Friedel–Crafts reactions | 109 |
| PIL-oxalic acid | 290 | 3-10 | 1.35 | Ionic complexation | 110 |
| P(CMVIm$Br_{1.03}$-*co*-AA) | 260 | 6-12 | 0.49 | Ionic complexation | 111 |
| MesoPIL | 260 | 6-20 | – | Ionic complexation | 112 |
| PIL-PAA | 330 | 2-50 | 1.1 | Ionic complexation | 36 |
| NP-imidazolium | 537 | 0.38 | 0.21 | Two-step polymerization | 155 |
| Meso-macroPIL-Br | 205 | several to tens of nanometers | 0.57 | Free radical homopolymerization | 103 |

**Abbreviations in Table 1**: PCP = porous cross-linked polymers; ABN = anionic borate network; ICOF = ionic covalent organic framework; PAF = porous aromatic framework; PP = phosphonium bearing polymer; CCLIL = cation cross-linked ionic liquid; beti = bis(pentafluoroethylsulfonyl)imide; $Tf_2N$ = bis(trifluoromethane sulfonyl)imide; NHC = N-heterocyclic carbine; T-IM = tubular microporous organic networks bearing imidazolium salts; EOF = element organic framework; POP = porous organic polymer; EB = ethidium bromide; BPP = Berkeley porous polymer-1; PCIF = POSS-based porous cationic framework; DVB = divinylbenzene; BVI$PF_6$ = 1-butyl-3-vinylimidazolium hexafluorophosphate; HCP = hyper-cross-linked porous polymers; CMIm = cyanomethyl vinylimidazolium; PIL = poly(ionic liquid); PAA= poly(acrylic acid); NP = nanoporous.

### 2.2.1 Synthesis of microporous ionic organic networks

In contrast to conventional synthesis of meso- and macroporous polymers, where long backbones formed by linear connection of polymerized monomers are interconnected by ditopic crosslinkers, microporous organic networks are typically constructed from monomer units that are multitopic (three or more connection points). It is known that smaller pores experience higher capillary pressure and higher surface energy, therefore the pores are instable and prone to collapse by bending and twisting of polymer chains to pack space efficiently.[6] In this regard, rigid and contorted polymer chains or high-degree connectivity/crosslinking enables polymers to maintain micropores and preclude/minimize swelling *via* adsorbing vapor or solvent molecules, which is advantageous over classically crosslinked polymers that are prone to pore swelling in contact with gas or solvent. However, it should not exaggerate the role of microporosity from an application viewpoint although it is a family of promising material that gains increased attention. Some excellent reviews on the scope of non-charged microporous polymers have been published, which are deserved to be consulted by readers who are interested in this area.[4, 12, 14, 15, 74-76]

Compared with neutral microporous networks, the synthesis of microporous ionic organic networks faces more challenges since the presence of ionic groups restricts the choice of solvents and available coupling chemistry. To achieve permanent porosity in microporous ionic organic networks, a crucial point is the choice of comparatively rigid monomers that, when crosslinked, yield pores with similarly rigid walls. Therefore, a rigid conjugated skeleton with reactive sites is a classic starting point. Similar to conventional neutral microporous materials, the principle for the synthesis of microporous ionic organic networks has drawn from an enormous number of modern bond-forming methodologies (*e.g.*, molten salt induced polymerization,[77,78] metal-catalyzed cross-coupling reactions,[79] hypercrosslinking reactions,[80] polycondensation reactions,[81] *etc.*) to yield a wide range of structural frameworks. The advantage of slowed-down bond-forming reactions is in favor of pores that more closely match the dimensions of potential guest molecules.[76] It should be noted that due to the difficulty to delicately control the pore structure, a small fraction of mesopores frequently accompany the prevailing micropores in the final porous product. Commonly the as-synthesized materials could be either amorphous or in an ordered crystalline state (*e.g.*, COFs), depending on the building blocks and the reaction mechanism. The geometry of the rigid building unit is highly relevant. Some typical building units of different configurations (such as planar triangle, octahedron, linear pattern, tetrahedron, and square, *etc.*) and symmetries (C2, C3, C4, and C6) and length developed for neutral microporous network formation are suitable as well for the construction of microporous ionic organic networks. Also for amorphous microporous ionic organic networks, although they are irregularly structured on larger scales, the local skeleton and nanoscale porosity of the materials can be still controlled by the initial monomers. In other words, the geometries of building units play vital roles in governing the structure of random networks. According to recent reports, microporous ionic organic networks could be synthesized from ionic monomers or through ionization reactions between neutral monomers during network formation. The two strategies will be discussed in detail next.

### 2.2.1.1 Microporous ionic organic networks from ionic building unit(s)

The synthesis of microporous ionic organic networks straightforward from the ionic monomers experienced rapid advance in the past few years. Many kinds of bond-formation reactions such as Sonogashira–Hagihara cross-coupling reaction, Suzuki cross-coupling reaction, Yamamoto type Ullmann cross-coupling reaction, Schiff base reaction, and Friedel–Crafts reaction have been employed to construct networks with permanent porosity. The ionic monomers containing $B^{-}$, $P^{+}$, $N^{+}$ or charged metal-organic units are usually involved in the ionic network formation. For example, copolymerization of ionic monomer $Li[B(C_6F_4Br)_4]$ with 1,3,5-triethynylbenzene *via* Sonogashira–Hagihara coupling yielded a microporous ionic organic network, and the chemical structure of the ionic material was proven by $^{13}C$, $^{11}B$, and $^{7}Li$ NMR spectroscopy.[40] The material showed a permanent porosity with $S_{BET}$ of 890 $m^2\ g^{-1}$ and a pore volume of 0.61 $cm^3\ g^{-1}$. In comparison, the use of uncharged tetrakis(4-bromophenyl)methane as building unit under similar polymerization condition yielded a material with $S_{BET}$ of 761 $m^2$ $g^{-1}$ and a pore volume of 0.54 $cm^3$ $g^{-1}$. The feasibility of direct employment of symmetric ionic building units for microporous ionic organic networks was further demonstrated by Long[82] and Zhu *et al.*[83] Similar ionic monomer derivatives (lithium tetrakis(4-bromo-2,3,5,6-tetrafluorophenyl)borate and lithium tetrakis(4-iodophenyl) borate)) were employed *via* polymerization with linear bis-alkyne linkers *via* Sonogashira–Hagihara polymerizations. The obtained materials gave $S_{BET}$ ranging from 82 to 480 $m^2$ $g^{-1}$.

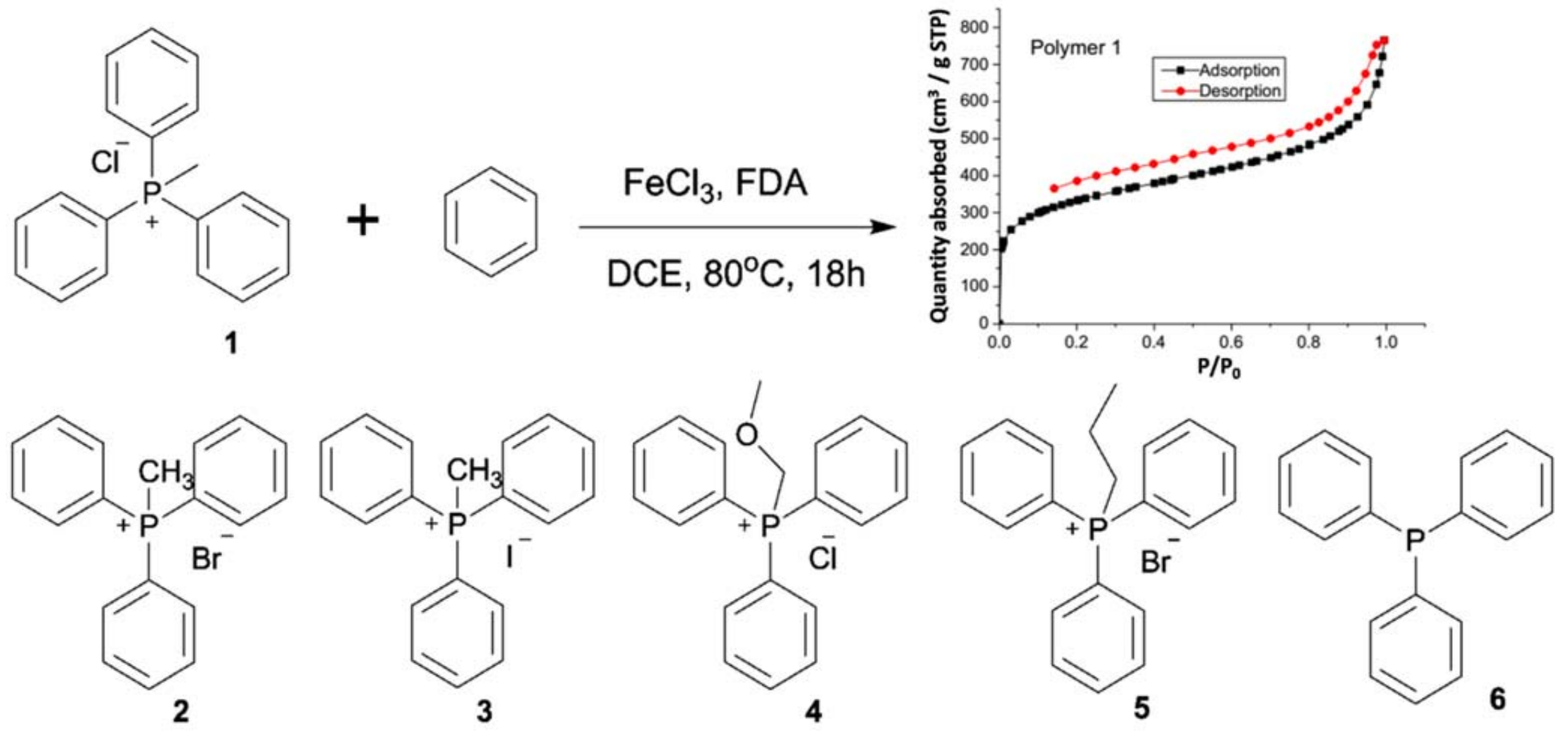

**Fig. 4** The synthesis of a series of hypercrosslinked phosphonium based microporous ionic organic networks (FDA = formaldehyde dimethyl acetal; DCE = dichloroethane). Reprinted with permission from ref. 85. Copyright 2015 Royal Society of Chemistry.

The report of phosphonium containing microporous ionic organic networks dates back to 2012.[84] The materials were synthesized by nickel(0)-catalyzed Yamamoto-type cross-coupling reaction with tetrakis(4-chlorophenyl)phosphonium bromide as the building unit. The resultant materials exhibited $S_{BET}$ of 650 to 980 $m^2$ $g^{-1}$ and average pore size of 0.7 to 1.4 nm on the basis of different counter anions. Zhang *et al.* developed a series of hypercrosslinked porous polymers with incorporated phosphonium salts *via* Friedel–Crafts reaction (Fig. 4).[85] Benzene was used as the co-monomer for the synthesis to avoid the limitation of Friedel–Crafts reaction for electro-deficient aromatic units. These porous materials have high $S_{BET}$ up to 1168 $m^2$ $g^{-1}$.

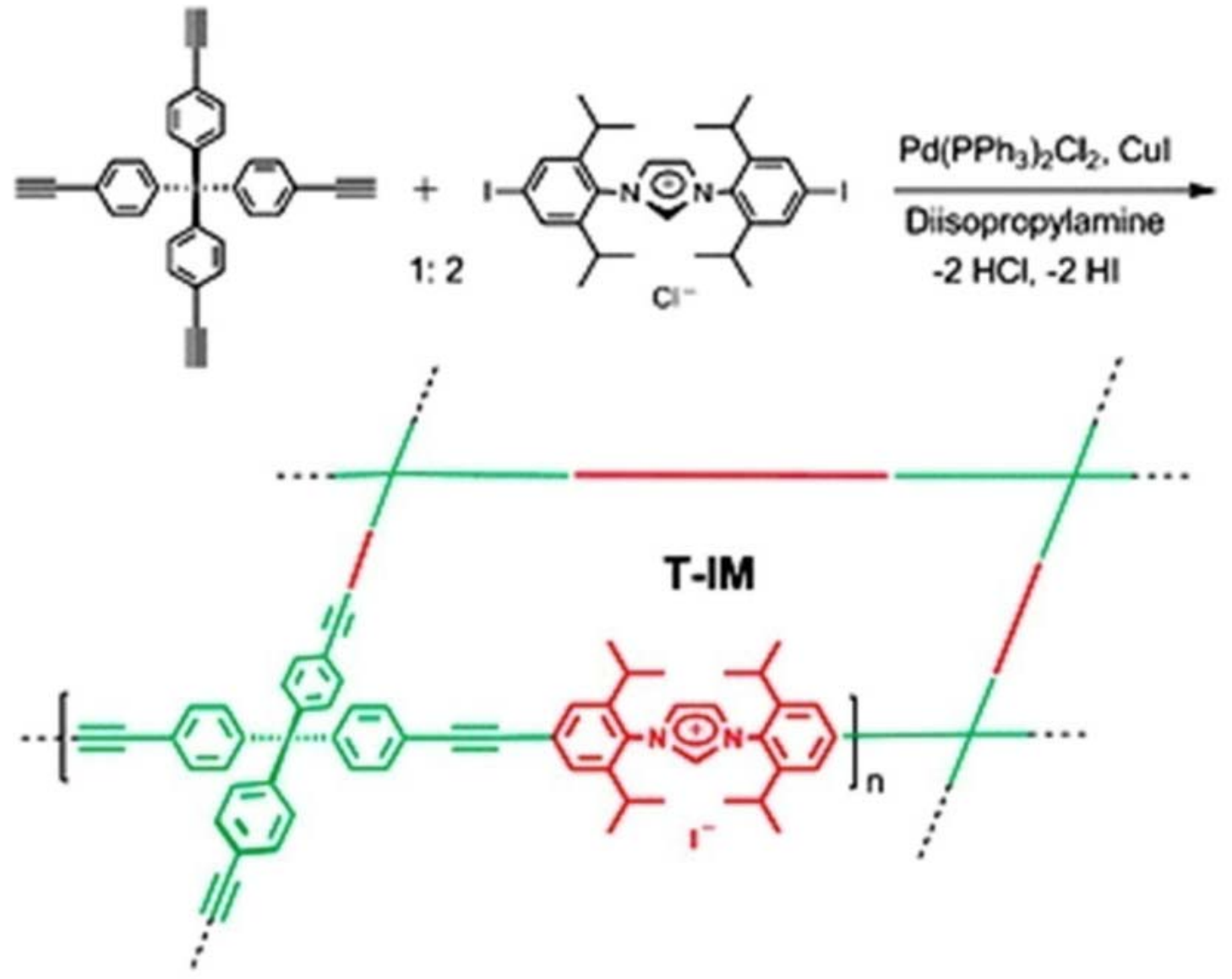

**Fig. 5** Preparation of porous organic networks bearing imidazolium salts by Sonogashira coupling reaction. Reprinted with permission from ref. 87. Copyright 2011 Royal Society of Chemistry.

One of the prototypical approaches toward synthesis of quaternary ammonium containing microporous ionic organic networks was based on polycondensation reactions. By appending a nitrile functionality onto the imidazolium unit, Dai and co-workers successfully developed a series of microporous ionic organic networks based on cation-crosslinkable ionic liquids.[86] The key structural feature of the IL for synthesis of porous polymers is the presence of the functional nitrile groups that can trigger crosslinking reactions at elevated temperatures, which underwent cyclotrimerization reactions between nitrile groups on neighbouring IL cations to give a dynamic amorphous polytriazine networks. The negligible vapour pressure and the nitrile group of these ILs were the most essential features to obtain porous materials under ambient pressure without catalyst or external template. The resultant materials showed tuneable $S_{BET}$ in a large range from 2 to 814 $m^2$ $g^{-1}$ when changing anions from $Cl^-$ to bis(pentafluoroethylsulfonyl)imide.

Another synthetic example of quaternary ammonium containing NIONs was reported by Son and co-workers (Fig. 5).[87] In their attempt, the tubular microporous organic network bearing imidazolium cations was prepared by Sonogashira coupling reaction of tetrakis(4-ethynylphenyl)methane and diiodoimidazolium salts. The resultant material offered $S_{BET}$ of 620 $m^2$ $g^{-1}$ and pore volume of 0.36 $cm^3$ $g^{-1}$. Besides, microporous ionic organic networks with $S_{BET}$ up to 262 $m^2$ $g^{-1}$ by incorporation of imidazolium linkers based on bifunctional aryl bromides *via* crosslinking with tetrafunctional boronic acids was reported by Kaskel, Glorius and co-workers.[88] In addition to imidazolium, a pyridinium unit is an equally interesting cation to be incorporated. Buyukcakir *et al.* reported bipyridinium-bearing porous network by Sonogashira coupling reaction between tetrakis(4-ethynylphenyl)methane and 1,1′-bis(4-iodophenyl)-4,4′-bipyridinium salts.[89] $S_{BET}$ could be effectively tuned in the range between 433 and 755 $m^2$ $g^{-1}$ by counter anion variation ($Cl^-$, $BF_4^-$ and $PF_6^-$). Similar ionic networks by using Sonogashira cross-coupling reaction between tris(p-ethynylphenyl)amine / 1,3,5-tris(4-ethynylphenyl)benzene and 1,1′-bis(4-bromophenyl)-4,4′-bipyridinium chloride with $S_{BET}$ up to 960 $m^2$ $g^{-1}$ were reported.[90] Very recently, a highly crystalline cationic COF with high thermal and chemical stability to combine a cationic monomer, ethidium bromide (EB) (3,8-diamino-5-ethyl-6-phenylphenanthridinium bromide) with 1,3,5-triformylphloroglucinol (TFP) by Schiff base reactions was reported (Fig. 6).[91] The $S_{BET}$ of resultant materials gradually dropped upon increase of the size of guest anions from $F^-$ (1002 $m^2$ $g^{-1}$) to $PW_{12}O_{40}^{3-}$ (8 $m^2$ $g^{-1}$) through anion exchange.

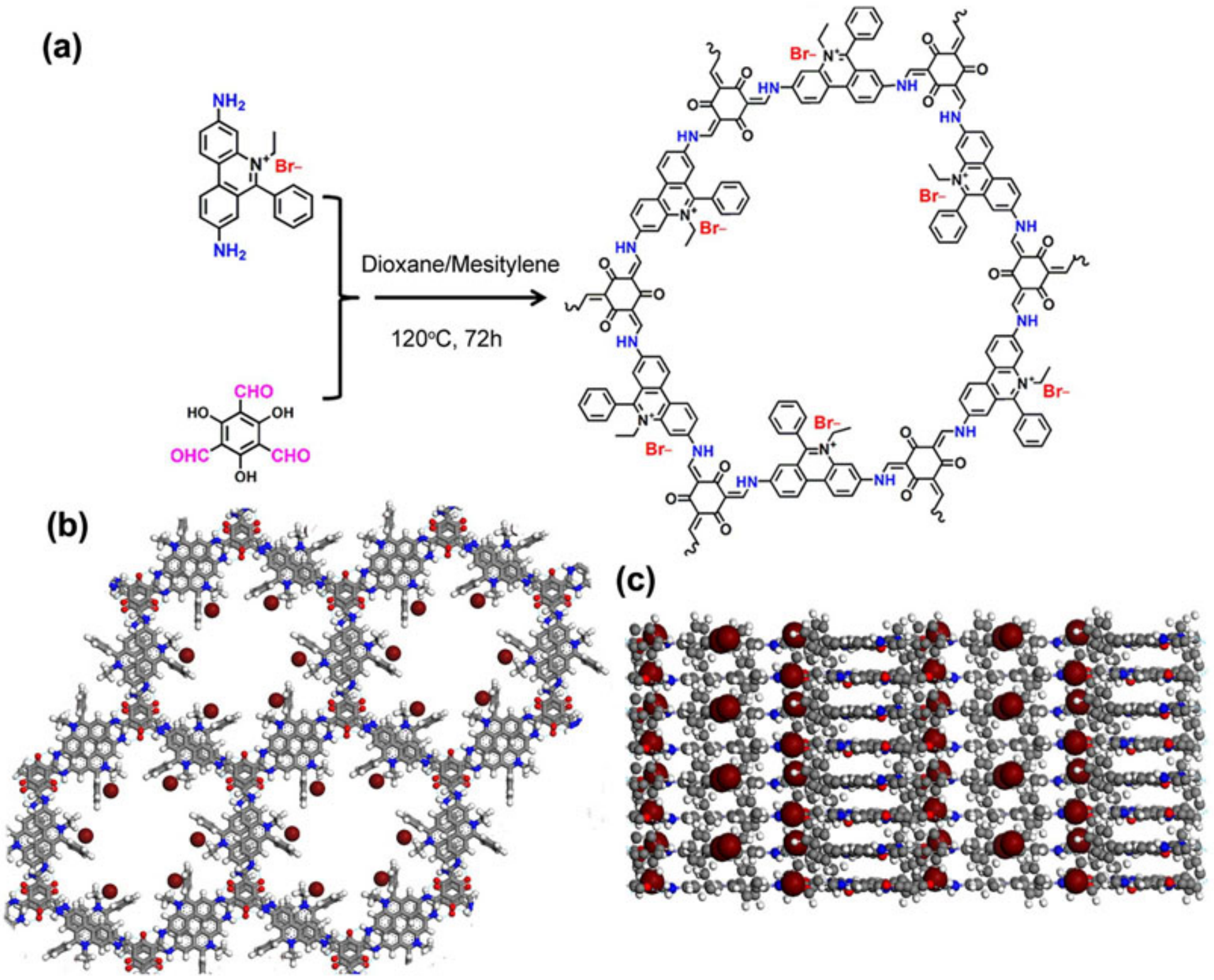

**Fig. 6** (a) Schematic representation of the synthesis of EB-COF:Br, (b) top views and (c) side views of the offset stacking structure of the EB-COF:Br. Reprinted with permission from ref. 91. Copyright 2016 American Chemical Society.

The direct incorporation of charged metal-organic units into polymeric skeleton is an alternative approach toward construction of microporous ionic organic networks. For example, Lin *et al.* introduced the charged building blocks $[Ru(bpy)_3]^{2+}$ and $[Ir(ppy)_2(bpy)]^{+}$ into porous networks by copolymerization with tetra(4-ethynylphenyl)methane *via* $Co_2(CO)_8$-mediated trimerization.[24] The produced materials have high surface areas with $S_{BET}$ of 1547 and 1348 $m^2$ $g^{-1}$ for Ir-NION and Ru-NION, respectively. Moreover, the materials exhibited remarkable chemical stabilities, *e.g.,* in concentrated hydrochloric acid, and were thermally stable in air up to 350 ℃. The incorporation of metal-N-heterocyclic carbenes (MHNCs) units into porous network was reported as well. The process generally involves the preparation of N-heterocyclic carbenes (NHCs) in the porous polymer backbone, and then the incorporation of metal ions into the polymer skeleton by reaction with the highly active carbene carbon. Based on the above-mentioned procedure,

Thomas *et al.* reported Cu(II)-coordinated polyNHC network, however with a relative poor surface area ($S_{BET}$: 62 m$^2$ g$^{-1}$).[92] Using longer alkyne-bearing building blocks, larger surface areas in an Au-NHC functionalized porous network with $S_{BET}$ up to 798 m$^2$ g$^{-1}$ were reported.[93] The combination of NHC–metal complex with hypercrosslinked polymer is an effective approach to prepare high surface area ionic network. For example, a Pd(II)-coordinated polyNHC hypercrosslinked network carried high surface area with $S_{BET}$ of 1229 m$^2$ g$^{-1}$.[94]

Besides, some other anionic units such as carboxylate [95] and phosphate [96, 97] groups have been incorporated into polymeric networks as well.

### 2.2.1.2 Microporous ionic organic networks from non-ionic building units

In addition to the direct use of ionic monomers, microporous ionic organic networks could also be synthesized *via* bond formation induced ionization from non-ionic monomers. One case is based on the nucleophilic substitution polycondensation. Chen *et al.* reported a series of polyhedral oligomeric silsesquioxane (POSS)[98] based porous cationic frameworks by polycondensation of octakis(chloromethyl)silsesquioxane (ClMePOSS) and rigid N-heterocyclic crosslinkers (4,4′-bipyridine analogues) *via* nucleophilic reactions.[99] The networks exhibited high surface areas ($S_{BET}$ up to 1025 m$^2$ g$^{-1}$) and large pore volumes (up to 0.90 cm$^3$ g$^{-1}$) due to the rigid POSS unit as connecting site in the network, which could fairly stabilize the skeleton and reduce the degree of interpenetration of network. The combination of nucleophilic substitution and self-condensation of 4-pyridinylboronic acid and cyanuric chloride by a one-pot method resulted in a quaternary pyridinium ion-decorated ionic network (PAF-50).[100] This material gave a moderate surface area with $S_{BET}$ of 532 m$^2$ g$^{-1}$. Another approach to microporous ionic organic networks is *via* the formation of a spiroborate linkage,[101] a kind of ionic derivative of boronic acid that can be formed readily through the condensation of polyols with alkali tetraborate, or boric acid, or through the transesterification between borate and polyols in a thermodynamic equilibrium. One example of ionic COFs by introduction of spiroborate linkage into the network exhibited a high surface area with $S_{BET}$ up to 1259 $m^2\ g^{-1}$ (Fig. 7).[50] It is stable in water as no obvious decrease of pore surface area was observed after 2 days of storage in water.

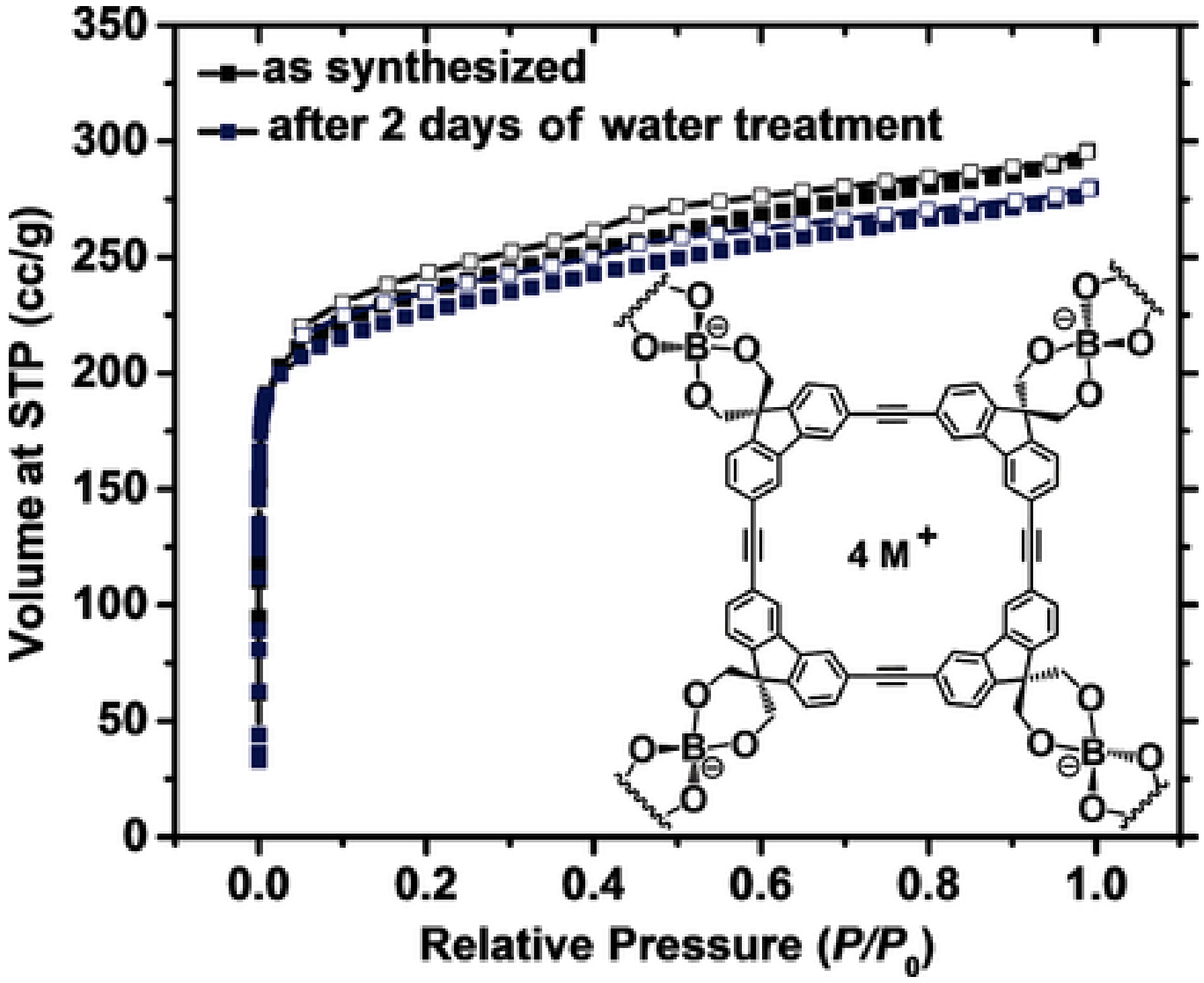

**Fig. 7** The ionic covalent organic framework (ICOF) containing $sp^3$ hybridized boron anionic centers was constructed by formation of spiroborate linkages and the corresponding sorption isotherms. Reprinted with permission from ref. 50. Copyright 2014 Wiley-VCH.

### 3.2.2 Direct free radical polymerization

Apart from polycondensation, the direct polymerization based on other mechanisms such as free radical polymerization is able to yield NIONs. In a typical procedure, a homogeneous mixture solution containing monomer, cross-linker, porogen molecules, and initiator is *in situ* polymerized to induce the phase separation between the cross-linked polymer chains and the porogenic solvent.[6] To prepare a porous network, the porogenic solvent employed here should be a relatively good one for the monomers, but a thermodynamically poor one for the resulting polymer networks. The obtained materials based on this method may contain micro-, meso- and macropores depending on the polymerization condition. A family of vinyl-containing ILs as precursors for NION synthesis by homopolymerization was reported. In this regard, using analogous nonpolymerizable ILs as porogenic solvents that are miscible with the ionic monomer is effective to generate NIONs with micropores and small mesopores. Odriozola *et al.* reported a simple, fast, sustainable, and scalable strategy to prepare NIONs based on radical homopolymerization of bisvinyl-functionalized ILs (*e.g.* a monomer prepared from 1-vinylimidazole and 1,4-dibromobutane) in the presence of the analogous nonpolymerizable IL 1-butyl-3-methylimidazolium bis(trifluoromethane sulfonyl)imide (Fig. 8).[102] In this case, the IL featuring a similar chemical nature as the PILs was regarded as an ideal solvent to obtain materials with small pores, as the IL and the *in situ* formed PILs were uniformly mixed rather than phase separated from each other during the polymerization. The IL used as porogenic solvent can be extracted easily after polymerization for reuse. Different monomer/porogen weight ratios (M/P = 1:0, 1:0.5, 1:1, 1:2, 1:3, 1:5, and 1:10) were systematically adopted to investigate their effect on the formation of the porous PILs. The optimized material (M/P = 1:1) produced uniform mesopores with pore size ranging from 6.8 to 8.5 nm and a $S_{BET}$ of 86.21 m$^2$ g$^{-1}$.

Another example of NIONs with shaped macroscopic morphology could be prepared through the free radical homopolymerization of a series of rigid bis-vinylimidazolium salt monomers.[103] The resultant meso-/macroporous monoliths exhibited $S_{BET}$ and pore volume up to 224 m$^2$ g$^{-1}$ and 0.57 cm$^3$ g$^{-1}$, respectively.

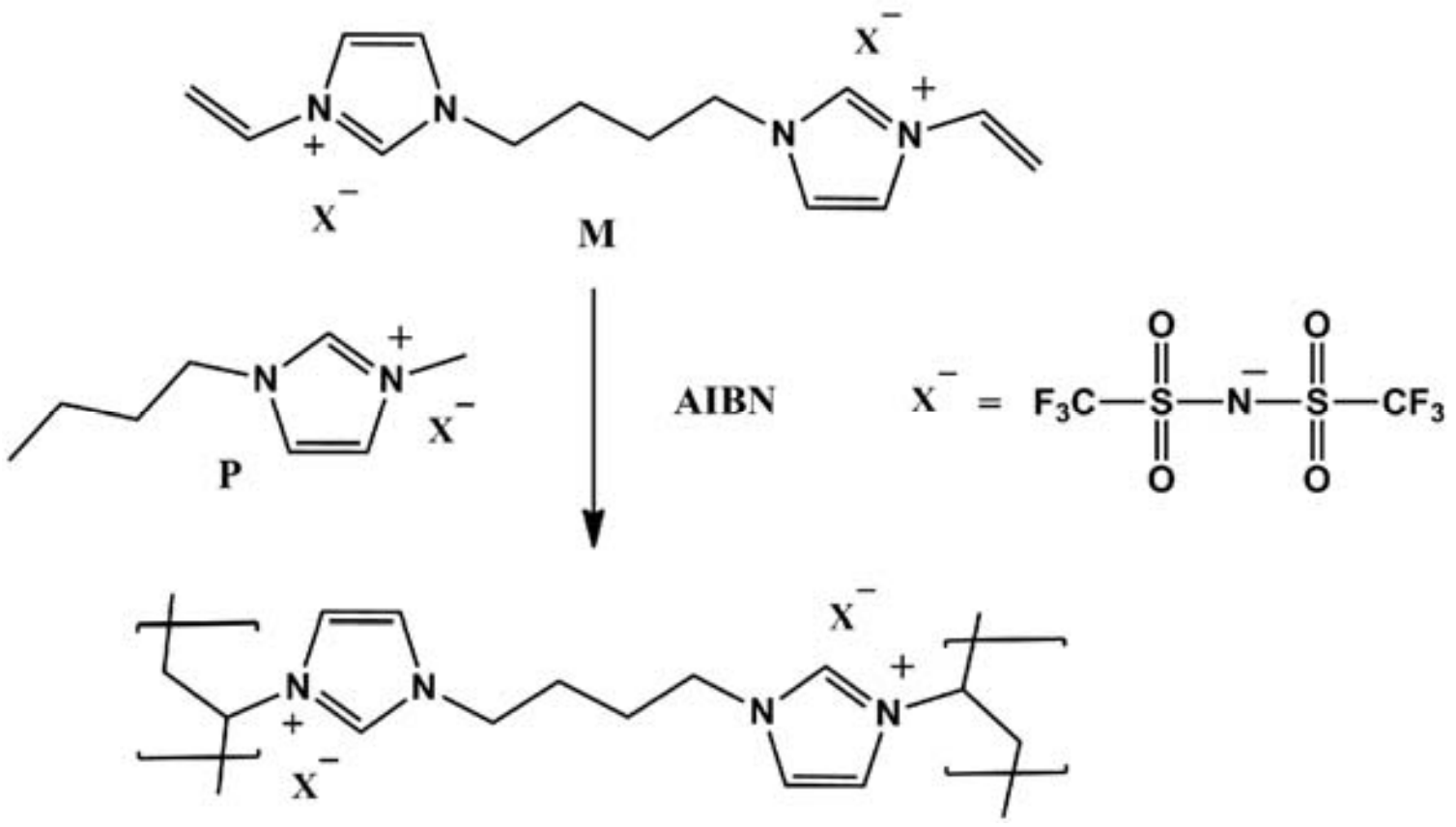

**Fig. 8** Synthetic pathway for PIL-based NIONs *via* a homo-polymerization approach. Reprinted with permission from ref. 102. Copyright 2014 Wiley-VCH.

Diverse NIONs have also been created by radical copolymerization. Wang, Zhou and co-workers reported mesoporous PILs synthesized by radical copolymerization of 1-aminoethyl-3-vinylimidazolium bromide with divinylbenzene plus the ion exchange of bromide anions with hydroxides.[104] $S_{BET}$ of the resultant materials goes up to 429 $m^2$ $g^{-1}$ with average pore size ranging from several to ten nanometers. The pore surface could also be functionalized with other substituents ($—C_4H_9$, —CN and $—NH_2$) *via* direct use of corresponding ionic monomers, which provided abundant opportunities to prepare advanced NIONs for task-specific applications.[105] Free radical copolymerization under hydro/solvothermal conditions is a simple and scalable method to fabricate NIONs. The hydro/solvothermal approach requires a closed environment in which the majority of the solvents remained in a liquid state at relatively high temperature and self-generated pressure. When ionic monomers started to polymerize under these conditions, a highly crosslinked network gradually forms in the solvent that serves simultaneously as a porogen. After the removal of solvents as the guest molecules, ionic polymers with open disordered porosity (generally in the mesopore range) are obtained. For example, direct copolymerization of divinylbenzene (DVB) and monomeric ILs with tuneable content generates mesoporous ionic networks under solvothermal conditions (Fig. 9).[106] This material offered high surface areas with $S_{BET}$ up to 935 $m^2$ $g^{-1}$ as well as large pore volume of 0.77 $cm^3$ $g^{-1}$, when using 1-butyl-3-vinylimidazolium hexafluorophosphate ($BVImPF_6$) as the IL monomer and THF as the solvent.

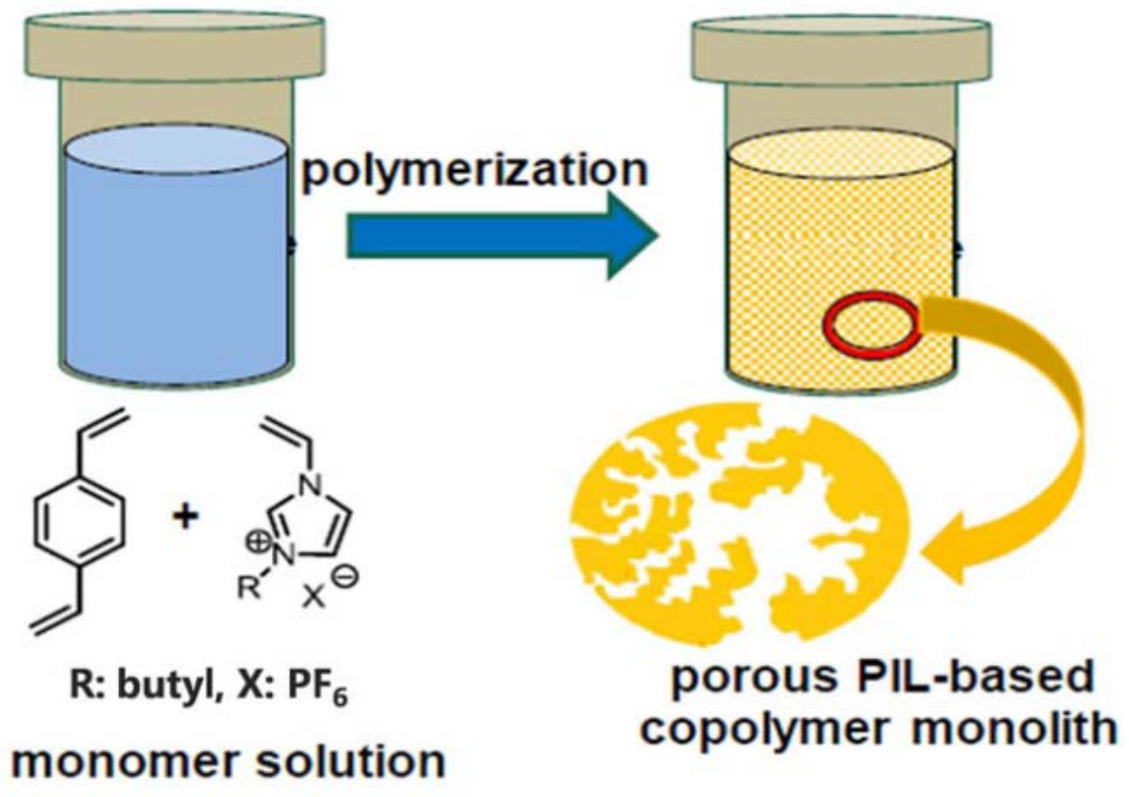

**Fig.9** The procedure for the solvothermal synthesis of NIONs by free radical copolymerization approach. Reprinted with permission from ref. 106. Copyright 2014 Elsevier.

It is notable that NIONs obtained by free radical polymerization method in some cases give relatively low surface areas in comparison with other methods. This is perhaps attributed to the difficulty in controlling the radical polymerization kinetics, leading to heterogeneities and skin layers. To improve this, the combination of radical copolymerization with other polymerization methods can be applied to prepare NIONs with larger surface area.[107] The main difficulty faced in the simultaneous conduction of two kinds of polymerization is the selection of an optimal reaction medium, as homogeneity is important for the formation of well-shaped hypercrosslinked structures. In 2005, Davankov *et al.* reported the synthesis of hydrophilic NIONs *via* combination of Friedel-Crafts alkylation reaction and free radical copolymerization of 4-vinylpyridine and *p*-xylylene dichloride in IL 1-butyl-3-methylimidazolium tetrafluoroborate (MBI-TFB).[108] The anion of the resultant porous PIL is the leaving group of the corresponding alkylation reagent. The fact that the hypercrosslinked structures were formed most reliably in this IL suggested that MBI-TFB was capable of supporting not only the alkylation but also polymerization reaction.

The use of such a combined method mentioned above could be extended to other systems. For example, a series of imidazolium based porous networks could be produced by using 4-vinylbenzyl chloride, divinylbenzene, and vinylimidazolium monomers with various alkyl groups (methyl, ethyl, or butyl) as precursors *via* combination of free radical copolymerization and Friedel-Crafts alkylation reactions. The final products exhibited $S_{BET}$ ranging from 447 to 667 $m^2\ g^{-1}$ with the pore volumes between 0.24 and 0.28 $cm^3\ g^{-1}$.[109] It was found that the incorporation of charge unit into the network did not significantly decrease $S_{BET}$, as compared with the hypercrosslinked polymer without imidazolium modified skeleton ($S_{BET}$: 728 $m^2\ g^{-1}$). This result unequivocally demonstrated that hypercrosslinked network can resist the pore collapse even when ionic species are introduced to the polymer porous structure.

### 2.2.3 Ionic complexation synthesis of NIONs

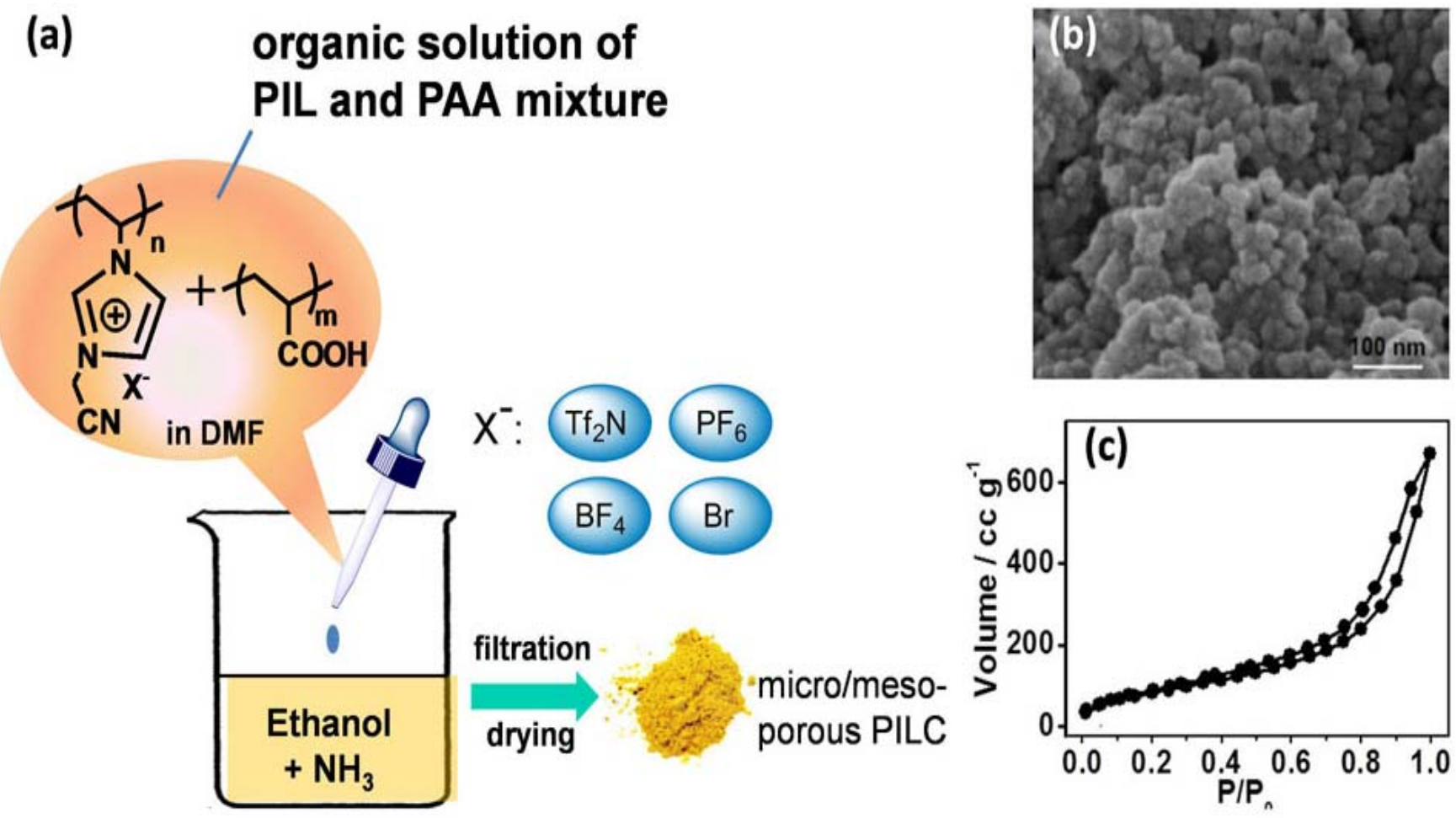

**Fig. 10** (a) Synthetic route to micro-/mesoporous PIL based on PCMVImX PILs and PAA [$Tf_2N$ = bis(trifluoromethane sulfonyl)imide]. (b) The corresponding SEM image of PCMVIm$Tf_2N$ nanoparticles and (c) its nitrogen sorption isotherms. Reprinted with permission from ref. 36. Copyright 2012 American Chemical Society.

Complementing traditional covalent bonded NIONs, recently our group has developed NIONs *via* interpolyelectrolyte complexation between a cationic PIL and a neutralized organic multi-acid/polyacid in organic media.[36, 110-112] Compared to the templating method involving template preparation and removal, the ionic complexation approach was simple in synthesis. The role of the organic multi-acid/polyacid is to balance the positively charged PIL and to crosslink it electrostatically into a stable porous framework. In 2012, the preparation of NIONs based on ionic complexation was first reported.[36] The work was exemplified with poly(3-cyanomethyl-1-vinylimidazolium) series (PCMVImX, X = $Tf_2N^-$, $PF_6^-$, $BF_4^-$, or $Br^-$ anions) and PAA (Fig. 10). As a typical procedure, PCMVIm$Tf_2N$ and PAA were dissolved in DMF to form a homogeneous solution because PAA in DMF was protonated and noncharged, thus no complexation occurred at this stage. This solution was added dropwise into an excess of ammonia-containing ethanol (denoted as complexation solvent) under sonication. Insoluble NION aggregates formed immediately *via in situ* ionic complexation between deprotonated PAA and the cationic PCMVImTf$_2$N. The $S_{BET}$ and the pore volume of obtained materials were measured up to 310 m$^2$ g$^{-1}$ and 0.98 cm$^3$ g$^{-1}$, respectively, which stemmed from the co-existence of micropores and mesopores. In the long history of polyelectrolyte research, interpolyelectrolyte complexation in aqueous phase has been intensively studied and used, while by contrast that in organic media has been rarely reported.[113,114] The reason is that conventional polyelectrolytes are dominantly water-soluble, the consequence of which is that most processing techniques of polyelectrolytes are water-based. Nevertheless, the efforts of interpolyelectrolyte complexation in aqueous phase, if without employing external inorganic template, so far ended up with non-porous or poorly porous products. PILs as a newly emerging class in the polyelectrolyte family have expanded the scope of polyelectrolyte complexation from aqueous into organic media due to their diverse and tuneable solubility in many organic solvents. The difference of complexation between in water and in organic solvents from a viewpoint of materials synthesis is that hydrophobic interaction is basically cancelled in organic media and hydrogen bonding can be weakened or cancelled as well, both of which are, besides the Coulombic interaction, crucial to define the microstructure of the complex products.

It is also noted that the proper choice of solvents to prepare the polymer mixture solution is essential, which should avoid deprotonation of PAA before adding into the alkaline solution for crosslinking. In fact, the fabrication concept was further expanded stepwise in terms of the building components, *i.e.* a PIL/organic multi-acid mixture, a PIL-*co*-PAA copolymer, and finally even a COOH-containing zwitterionic PIL homopolymer were reported, always adding on the possibilities.[110-112] The resultant materials containing micro-/mesopores showed potential applications in gas capture and separation, metal nanoparticle (MNP) immobilization and solvent purification.

Through modification of the fabrication procedure mentioned above, the products may include also nanoporous PIL membranes bearing a unique gradient of the degree of electrostatic complexation along the membrane cross-section.[43, 115-118] Commonly, the membrane fabrication was conducted according to the following procedure.[115] The PCMVImTf$_2$N and PAA mixture in DMF solvent was cast onto a glass plate and dried to produce a yellowish sticky thin polymer blend film. Then the film on the glass was immersed in an aqueous $NH_3$ or NaOH solution to induce *in situ* ionic complexation between PAA and the surrounding PCMVImTf$_2$N chains to build up the electrostatically crosslinked porous network (Fig. 11). Interestingly, the different gradient of crosslinking density can produce membranes containing different kinds of pores, and the resultant structure depended on the kind of organic acids. The chemical structure, such as the anion type and backbone architecture, of PILs can also modulate the nanopore size.

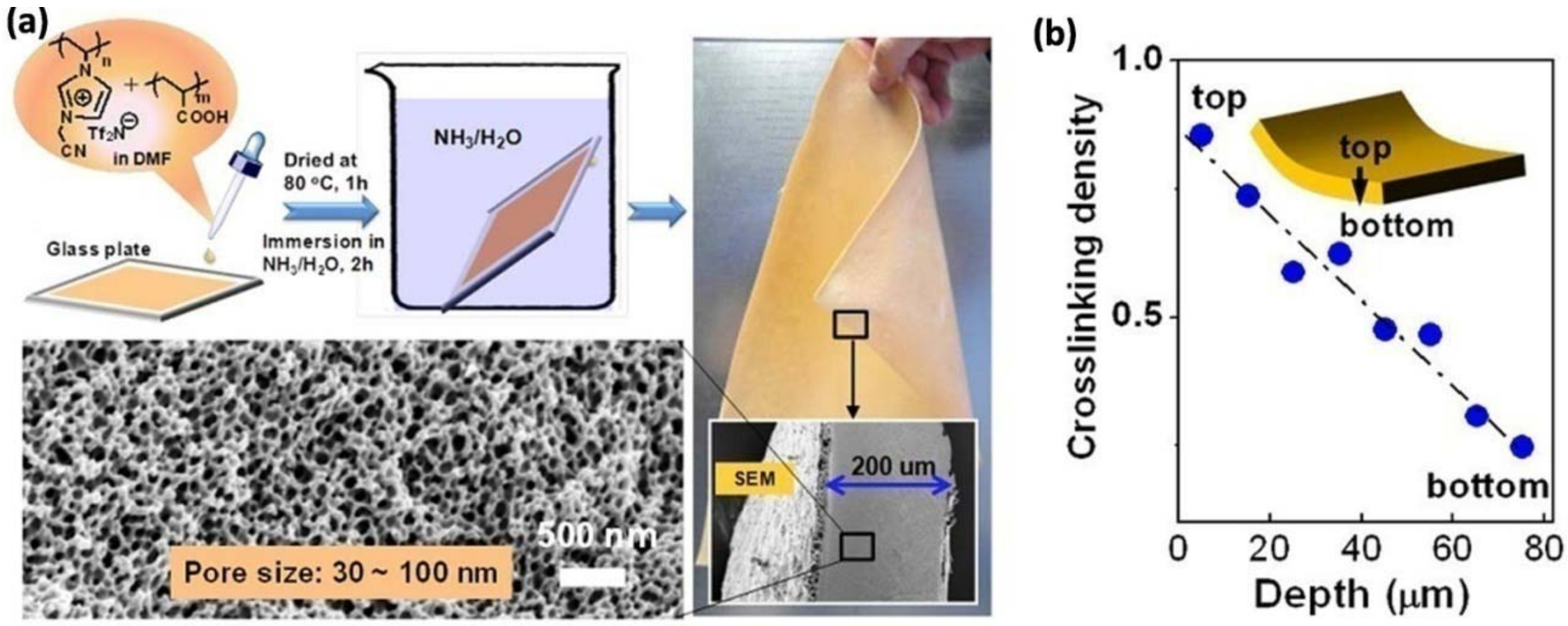

**Fig. 11** (a) Fabrication scheme of nanoporous polyelectrolyte membrane *via* an ionic complexation route using a PIL and PAA as the two structural components. (b) The figure depicts the crosslinking density profile along the diffusion tract of aqueous ammonia into the film. Reprinted with permission from ref. 115. Copyright 2013 American Chemical Society.

Moreover, such ionic complexation based porous membrane could be functionalized with other groups to develop advanced membranes. An interesting example of stimuli responsive porous membranes was reported by Vansco and co-workers (Fig. 12).[119] PIL containing the redox active species poly(ferrocenylsilane) (PFS) and PAA was combined by ionic complexation method to fabricate a porous membrane. A three-dimensionally interconnected porous structure with pore sizes of 70-250 nm was observed in that membrane. Interestingly, the pore of membrane showed responsiveness to external redox triggering. SEM images revealed that the membranes had a visibly higher density of the openings in an oxidized form, while it had a higher density of closed cells in a reduced state. Such redox responsive variation of the cellular morphology was attributed to the oxidation of the ferrocenyl groups of PFS. In fact, the switching process between open and closed porous states can be also triggered reversibly *via* chemical oxidation and reduction, *e.g.* by treating the membrane with 10 mM $Fe(ClO_4)_3$ and 10 mM ascorbic acid aqueous solution, respectively. Such switching behavior was further demonstrated to be used for tuning the pore permeability with average water flow rate varies between 0.092 ± 0.004 mL $cm^{-2}$ $s^{-1}$ (oxidized state) and 0.064 ± 0.005 mL $cm^{-2}$ $s^{-1}$ (reduced state).

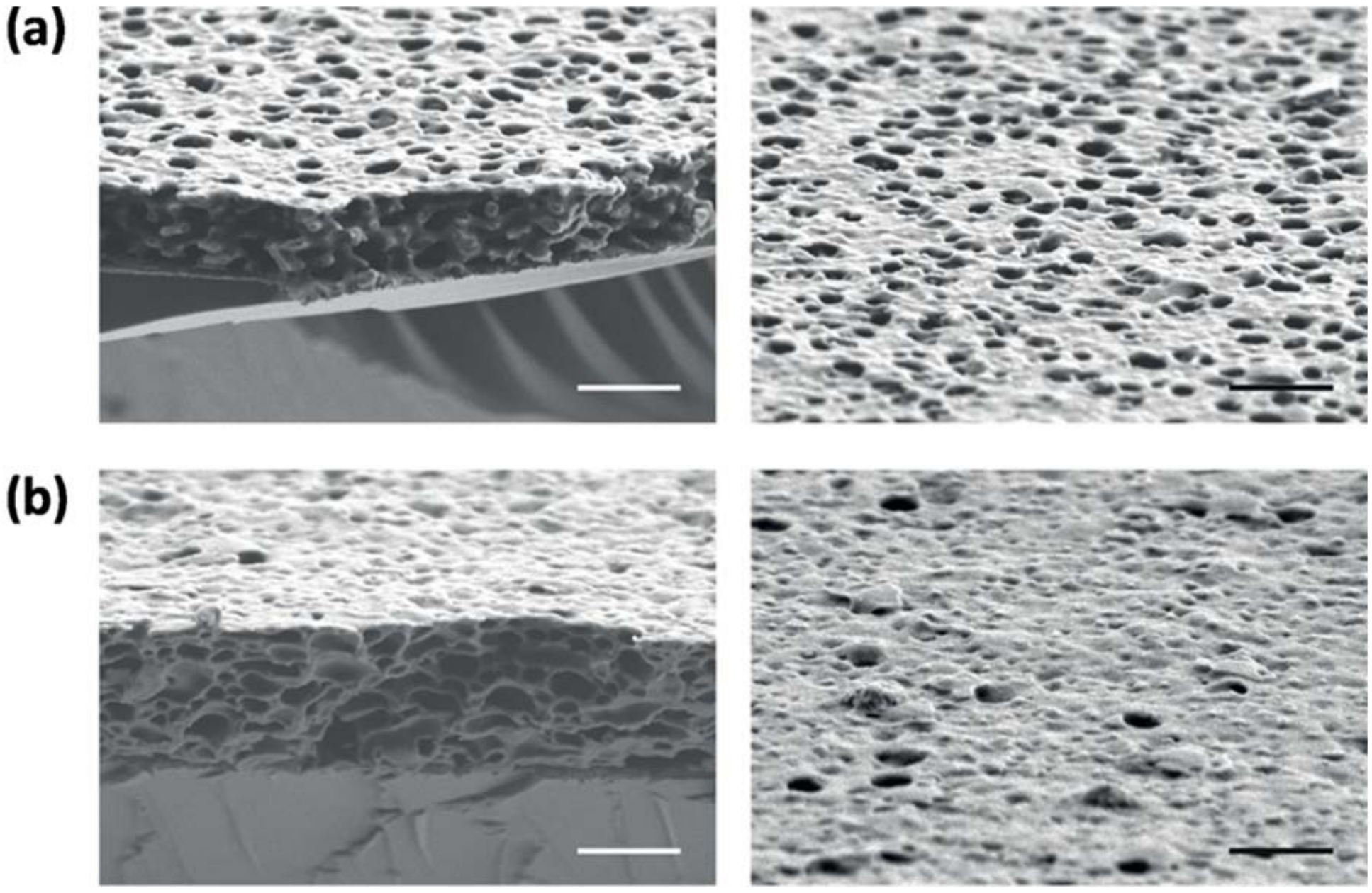

**Fig. 12** Cross-sectional and surface SEM images of porous membranes after oxidation at 0.6 V for 10 min (a), and then reduction at −0.2 V for 10 min (b). Scale bar: 1µm. Reprinted with permission from ref. 119. Copyright 2014 Wiley-VCH.

### 3.3 NIONs by post-synthesis

Besides the direct synthesis with or without template, post-synthetic modification (PSM) of neutral frameworks can be involved to generate NIONs. PSM has been recognized as an effective, versatile methodology, which allows bringing a variety of functional groups into the prefabricated network. It is especially true for NIONs that are difficult to achieve through one-step network synthesis discussed in the previous sections. A large number of existent porous materials, such as porous carbons, zeolites, mesoporous silicas, and MOFs have been modified by such method for functionalization. PSM serves as a platform to control over both the substituent and the degree of modification of parent porous structures, which can introduce multiple substituents stepwise or simultaneously into a single porous network, thus enabling fine-tuning and optimization of properties and functions of the resultant porous network. It must be noted that the PSM approach usually has an intrinsic negative impact on the surface area that in extreme can lead to very low porosity and a compromised surface area. One possible strategy to tackle this issue is to judiciously select porous polymer template with high surface areas and chemical stability as starting materials. In this regard, sufficient surface areas can be retained after the introduction of functional moieties onto the pore wall.

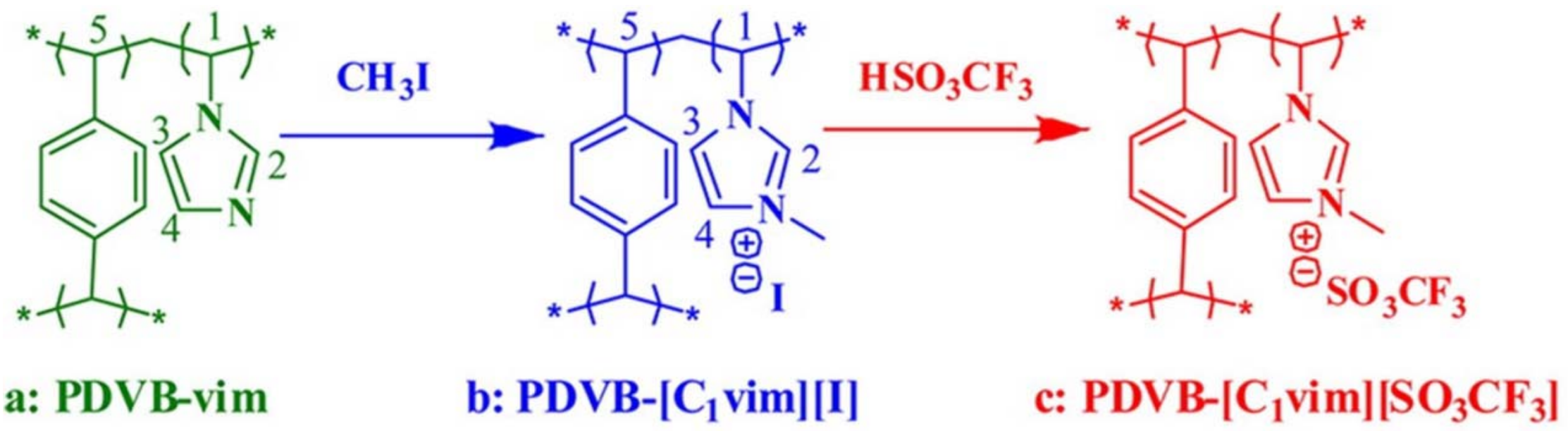

**Fig. 13** Scheme for the synthesis of PDVB-[$C_1$vim][$SO_3CF_3$] from PDVB-vim *via* post-quaternization of skeleton of organic skeleton network followed by anion exchange process. Reprinted with permission from ref. 41. Copyright 2012 American Chemical Society.

Cationization of a neutral polymeric network is popular to achieve NIONs by quaternization of network skeleton. For example, Liu *et al.* reported the synthesis of NIONs with pore size of 10–40 nm by quaternization (Fig. 13).[41] In their work, a neutral mesoporous polymer was first synthesized by solvothermal copolymerization of divinylbenzene (DVB) with 1-vinylimidazole (vim) in ethyl acetate as solvent at 100 °C in an autoclave. Through the quaternary ammonization of the prefabricated mesoporous polymers with $CH_3I$ and then anion-exchange treatment to replace iodide with trifluorosulfonate, the resultant PDVB-[$C_1$vim][$SO_3CF_3$] maintained the porosity of neutral counterpart with $S_{BET}$ and pore volume of 181 $m^2$ $g^{-1}$ and 0.75 $cm^3$ $g^{-1}$, respectively. Follow-up works based on a similar synthetic strategy with modified synthetic conditions from solvothermal to solution or solid-solution phase have been reported.[120-124]

Cationization of neutral polymeric networks could also be achieved through external stimuli.[125] For example, upon UV-light exposure, spiropyran containing porous polymer membranes underwent isomerization into the ionic merocyanine structure with charge-modified pore wall;[126] the redox-responsive polymer membrane containing poly(3-carbamoyl-1-(*p*-vinylbenzyl) pyridinium chloride) (PCVPC) can reversibly deionize the pyridinium cation by reduction or oxidation.[127]

Ionization of neutral networks by negative charges proceeds by post-introducing functional groups with different acidity or polarity, *e.g.*, –OH,[128] –$SO_3H$[129,130] –$PO_3H$[131,132] as well as –COOH[133,134]. In a typical run, the nanoporous network is directly immersed into the corresponding acid/base solution to conduct the surface modification. The resultant acid/base groups are easily protonated/deprotonated to generate the ionic network. The functional group together with charged characteristic leads to the resultant NIONs featuring target-specific applications such as solid acid/base catalysts or absorbents of environmental pollutants. Wu *et al.* reported a different method to functionalize polymer materials with well-ordered mesopores with –$SO_3H$. In their work, the ordered mesoporous polymers of FDU-14 and FDU-15 were first synthesized by a soft template method with $S_{BET}$ of 545 and 463 $m^2$ $g^{-1}$, respectively.[21] The sulfonation of the samples was then carried out in a Teflon beaker placed in a Teflon-lined autoclave *via* a gas-solid reaction (Fig. 14). The surface area and pore size distribution analysis of post-synthetic NIONs (FDU-14-$SO_3H$, FDU-15-$SO_3H$) revealed that the porosity of materials was well maintained without obvious damage ($S_{BET}$ for FDU-14-$SO_3H$ and FDU-15-$SO_3H$ were 539 and 447 $m^2$ $g^{-1}$, respectively.).

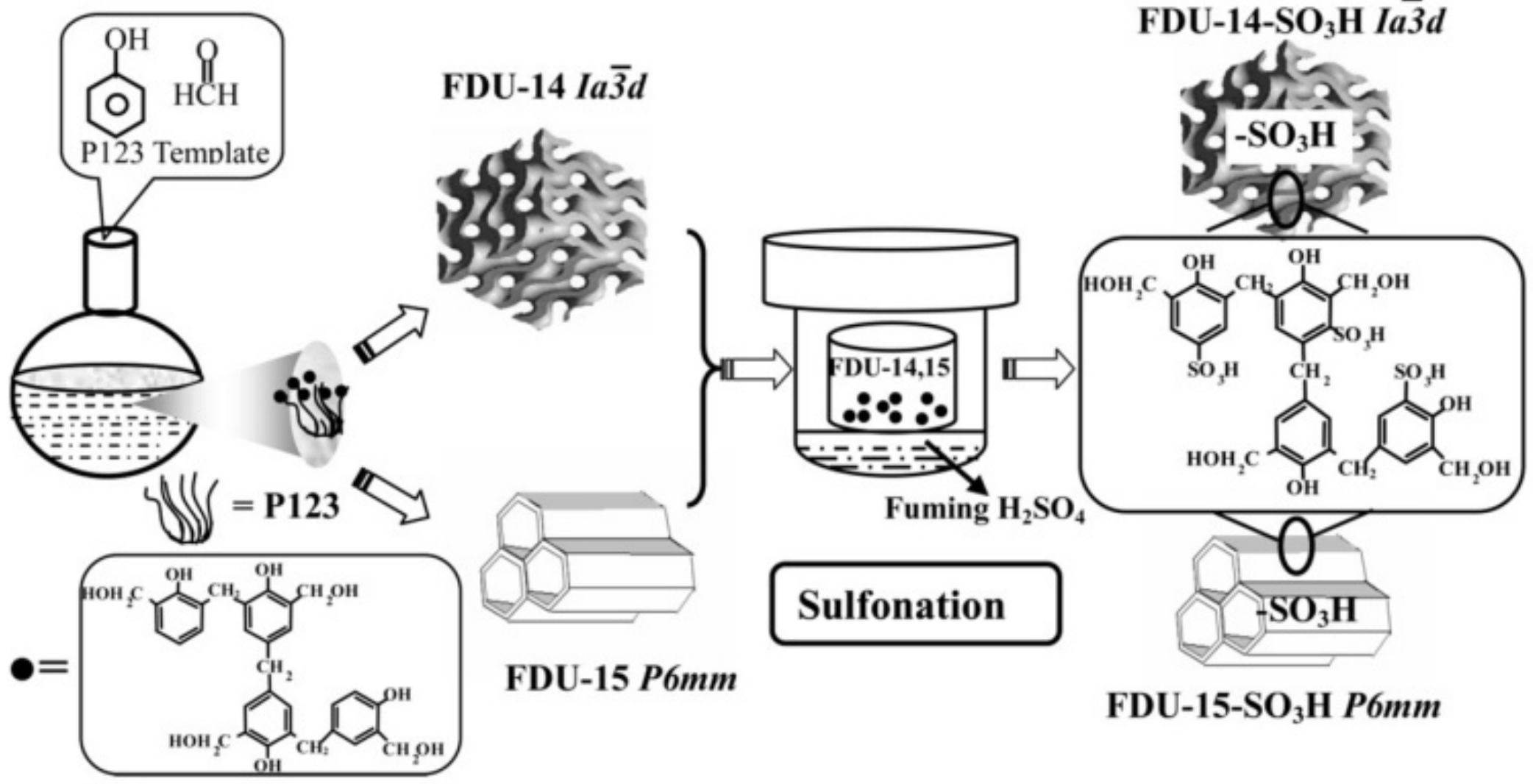

**Fig. 14** Schematic illustration for the preparation of $SO_3H$-functionalized mesoporous polymers with different mesostructures *via* a gas-solid reaction. Reprinted with permission from ref. 21. Copyright 2007 Wiley-VCH.

In addition, the ionization of porous polymeric network could be conducted by post-introducing metal ions into the skeleton through neutralization reaction.[135-140] It should be mentioned that the method presented here does not change the connection of prefabricated organic networks, while the introduction of metal ions only plays a role of modification of the network skeleton. A rigid scaffold is generally required to stabilize the network during this process without seriously degrading the porosity. For example, Zhu and co-workers reported a porous aromatic framework PAF-26-COOH featuring carboxyl-decorated pores, which can effectively bind light metal ions such as $Li^+$, $Na^+$, $K^+$ and $Mg^{2+}$ *via* a neutralization approach. The resultant metal-modified pore wall exhibited a moderate surface area without serious degradation of the pore structure.[136] There are also other works by introducing charged organometallic units (*e.g.*, $[M(bpy)_3]^{n+}$, M = Ru, Ir) onto the polymer skeleton, which enable, *e.g.*, light harvesting and catalysis.[24, 139-142] For example, Cooper *et al.* developed microporous conjugated organic polymers by post-loading metal ions into bipyridine-functionalized porous precursor. The resultant materials maintained high surface areas after modification.[140]

## 2. 4 Other methods

There are also other more exotic methods for the preparation of NIONs. In 2007, Texter *et al.* reported a synthetic strategy based on the microemulsion polymerization using a polymerized IL, 1-(2-acryloyloxyundecyl)-3-methylimidazolium tetrafluoroborate as monomer. The resultant material in an ionic gel form can be transformed into nanoporous polymers with pore size around 100 nm at a higher content of crosslinker (ethylene glycol dimethacrylate, EGDMA) up to 10 wt%.[143]

Recently, Zhang *et al.* reported a green, fast and efficient mechanochemical strategy to synthesize NIONs without template.[144] The chemistry here involved the quick formation of C-C and C-X (X=halide) bonds *via* a cross-coupling process.[145] Mechanochemical grinding of raw materials of 5,5,6,6-tetrahydroxy-3,3,3,3-tetramethyl-1,1′-spirobisindane and an IL (1-methyl-3-(2,3,4,5,6-pentafluorobenzyl)-imidazolium hexafluorophosphate) in the presence of a catalytic amount of $Pd(acac)_2$ yielded the desired NION. The resultant material carried hierarchical micro- and mesopores with $S_{BET}$ of 87 $m^2\ g^{-1}$. The $S_{BET}$ could be further improved up to 210 $m^2\ g^{-1}$ by post-modification with $Li^+$ ion.

## 4. Applications of NIONs

### 3.1 Catalytic applications

The use of recyclable catalysts for organic synthesis to minimize waste production and optimize catalyst efficiency is one of current goals for the pursuit of greener, safer, and more environment-friendly technologies in chemical and pharmaceutical industries.[146-150] NIONs with charge decorated pore surface and intrinsic porosity in different ranges of sizes present new opportunities for heterogeneous catalysts. Their use in catalytic applications will be reviewed in the following subsections according to the types of functional catalytic species.

#### 3.1.1 NIONs with functional organic group for catalysis

NIONs with intrinsic ionic units could be used as functional sites for catalytic application. One of important materials is the imidazolium-containing nanoporous polymers. In these catalysts, N-heterocyclic carbenes (NHCs) can be formed by *in situ* deprotonation of the C2 carbon in the cationic ring, thus constituting a class of very important organocatalysts.[151] The immobilization of a bifunctional imidazolium entity in a series of highly crosslinked organic framework materials was obtained by Suzuki coupling with tetrafunctional boronic acid linkers.[88] The resulting porous materials were applied as heterogeneous organocatalysts in the NHC-catalysed conjugated umpolung of α, β-unsaturated cinnamaldehyde. The product yield and the stereoselectivity reach up to 86% and 59 %, respectively, which are comparable to that of the molecular analogues (85% and 65%, respectively). Given that NHCs are known to activate epoxides for the reaction with $CO_2$ to produce cyclic carbonates, the NHC-based NIONs were extensively investigated in this model reaction owing to the economic and environmental benefits arising from the direct utilization of $CO_2$. In 2007, Han *et al.* firstly reported the employment of imidazolium containing NIONs for this catalytic experiment. The catalytic performance for the cycloaddition of $CO_2$ to epoxides was investigated.[152] It was demonstrated that the NION-based heterogeneous catalysts were very active with the yield up to 97.4%, stable during operation, and could be easily separated from the products for recycle use. After this finding, numerous works were reported by using NIONs with different pore structures for this kind of catalytic processes.[87, 89, 92, 153, 154] For the effective utilization of $CO_2$ as sustainable source, the fixation efficiency is one of the key parameters. One approach is to incorporate sterically confined NHCs into material skeleton, which impart stability to NHCs and preclude abstraction of acidic protons and intermolecular dimerization of NHCs. Recently, Coskun and co-workers reported a class of ionic microporous polymers by combination of advantages of micropores and sterically confined NHCs.[155] The material exhibited exceptional $CO_2$ capture fixation efficiency of 97% at room temperature, which is so far the highest value reported for carbene based materials measured in the solid state. The catalyst was tested in the conversion of $CO_2$ into cyclic carbonates at atmospheric pressure with excellent yields up to 98% along with 100% product selectivity. This material also showed substrate selectivity for the corresponding epoxides due to the narrow pore size distribution of the dominant micropores.

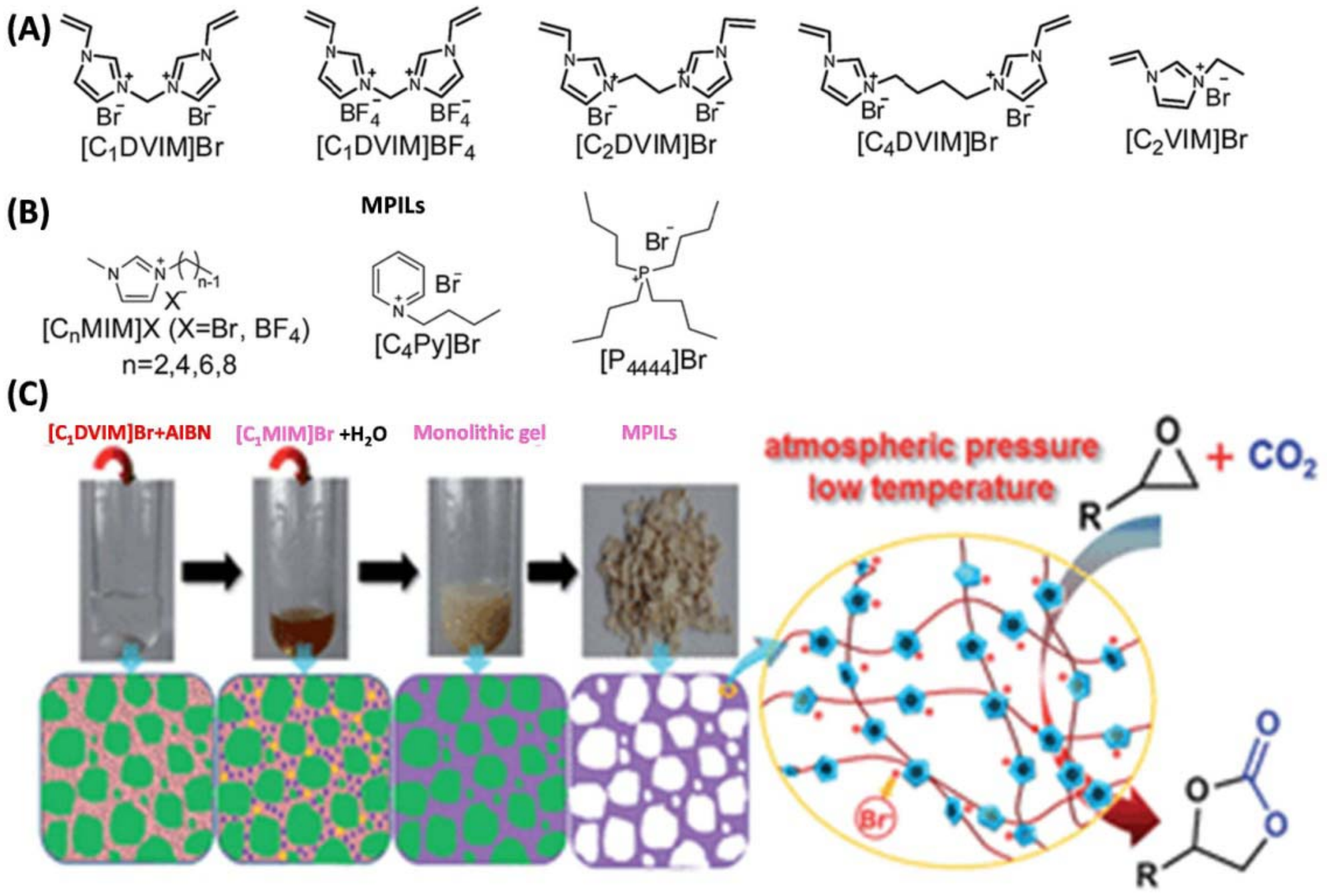

**Fig. 15** (A) Structure of the synthesized bis-vinylimidazolium salt monomers and (B) structures of the IL solvents. (C) Photographs of the ionothermal synthesis of a mesoporous PIL with schematic illustration of the possible mechanism for the formation of the mesoporous structure. The charged network acts as an efficient catalyst toward cycloaddition of $CO_2$ to epoxides. Reprinted with permission from ref. 103. Copyright 2015 Royal Society of Chemistry.

Although a number of materials enabling the activation of epoxides for the reaction with $CO_2$ to cyclic carbonates have been reported, pursuing the high reaction activity at mild conditions, *e.g.*, atmospheric pressure and low temperature, is still challenging. Recently, Wang and co-workers found that ionothermal-derived meso-/macroporous PILs from bis-vinylimidazolium salt precursors exhibited high activity toward efficient conversion of $CO_2$ at atmospheric pressure and low temperature among the reported metal–/solvent–/additive-free heterogeneous catalysts (Fig. 15).[103] Particularly, the as-synthesized NION catalyst exhibited a high yield (up to 90%) at lower temperature down to 70 °C under organic solvent-free conditions toward various substrates. Especially the inert long carbon-chain alkyl epoxides could be converted by the catalysts into corresponding cyclic carbonates with high yields (89 ~ 99 %) under ambient conditions (90 ~ 120 °C). Mechanism investigation revealed that the specific characteristic of NIONs with the large surface area and hierarchical meso-/macroporous structure enabled the good dispersion of active sites and accelerated the mass transfer of substrate molecules and products. Moreover, the intrinsic $CO_2$-philicity (arising from the $CO_2$-imidazolium interaction in the large-surface area pore structure) improved the local concentration of $CO_2$ around the catalytic centres inside the pores of the polymeric framework.

The NIONs functionalized with both acid/base and ionic groups could be used as solid acid/base catalysts. Solid acid/base catalysis has received much attention because of their potential applications for fine chemical syntheses with the advantages of easy separation of catalyst from the liquid reaction medium, reduced corrosion due to localized acids on the substrate rather than soluble ones, good recyclability, green chemical processes, and enhanced product selectivity. Mesoporous ionic organic network based solid acid/base catalysts are favourable because their high surface areas and large pore openings accelerate mass transfer, allowing bulky substrates to access the active sites embedded in the porous matrix. Xiao's group has conducted a series of work using sulfonate grafted porous PILs as solid acid catalysts.[156-158] These catalysts exhibited excellent catalytic activity toward various reactions such as esterification of acetic acid with cyclohexanol and 1-butanol, condensation of benzaldehyde with ethylene glycol as well as conversion of fructose to 5-hydroxymethylfurfural. Especially, they found that PDVB-[$C_1$vim][$SO_3CF_3$] with hydrophobic mesoporosity exhibited excellent activities in a series of catalytic reactions such as transesterification, the Peckmann reaction, Kharasch addition, esterification and hydration, normally with yields above 96%.

Asymmetric organocatalysis is an important segment in catalysis. To date, the use of NIONs as catalysts for heterogeneous asymmetric catalysis has been rarely explored.[93,94] One example of NIONs bearing 1,1′-binaphthalene-2,2′-diol (binol)-derived phosphoric acid chloride backbone for such kind of reaction was reported by Kundu and co-workers (Fig. 16).[97] It showed high activity and high enantioselectivity (*ee*) in the transfer hydrogenation of dihydro-2H-benzoxazine (conversion: 99%, *ee*: 98%) and 2-arylquinolines (conversion: 99%, *ee*: 98%), asymmetric Friedel–Crafts alkylation of pyrrole (conversion: 91%, *ee*: 96%) and an aza–ene-type reaction (conversion: 82%, *ee*: 81%). The catalytic activities were comparable to those of the corresponding soluble homogeneous catalyst. Moreover, the NION-based catalyst was stable, easily separable, and could be reused up to 10 times with constant conversions of 99 % and 98 % enantiomeric excess toward asymmetric dihydro-2H-benzoxazine reaction.

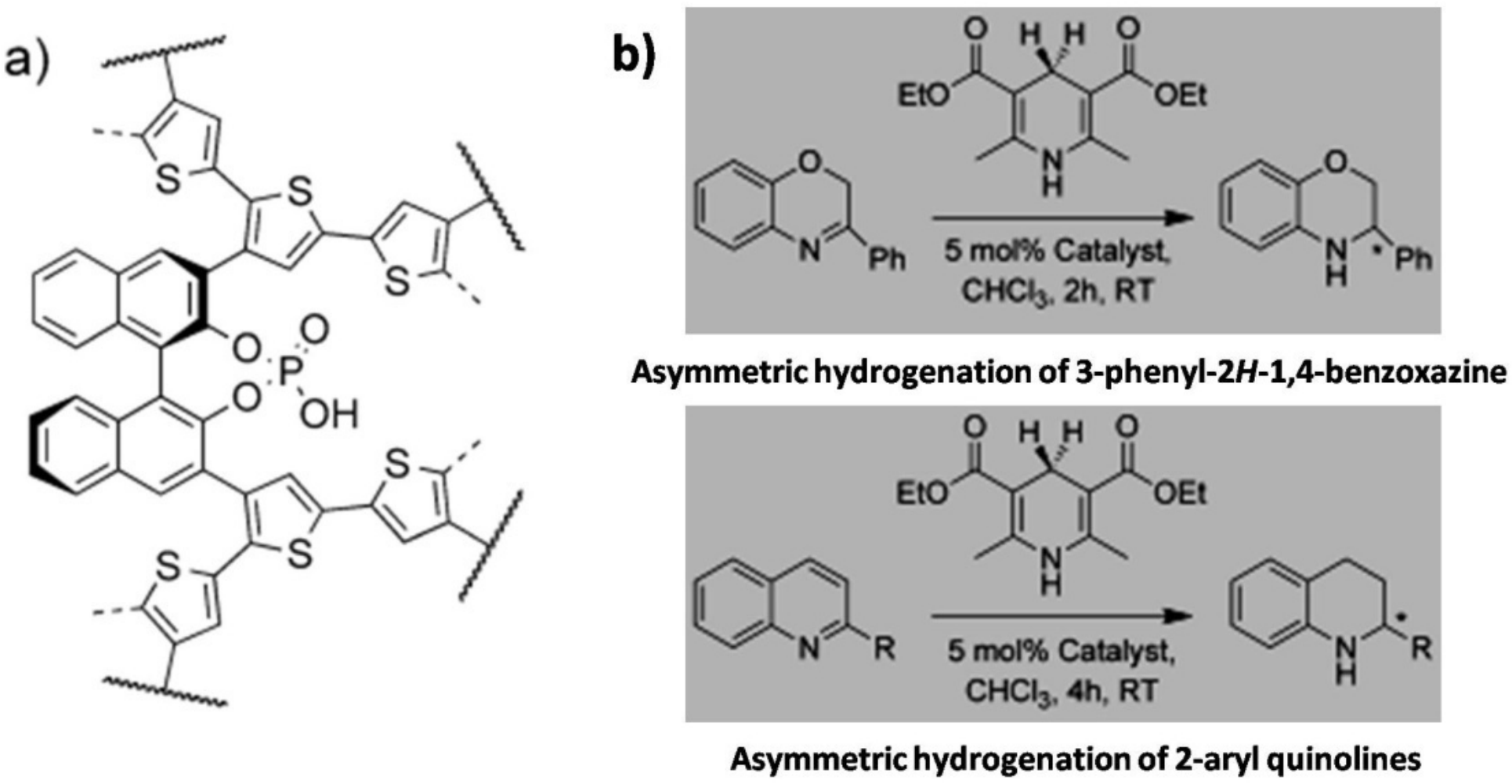

**Fig. 16** Example of asymmetric organocatalytic conversions (b) based on a chiral microporous polymer bearing the unit of binaphthyl phosphoric acid (a). Reprinted with permission from ref. 97. Copyright 2012 Wiley-VCH.

Although heterogeneous catalysts have activities similar to homogeneous base catalysts, their active sites are more sensitive and can be poisoned by molecules such as $H_2O$, $CO_2$, and fatty acids (FFAs), for instance by treating crude vegetable oils. To effectively address these problems, tuning the surface characteristic of NIONs is quite relevant. The hydrophobic surface is generally favourable for the development of solid base catalyst, which may otherwise suffer from moisture contamination. For example, copolymerization of divinylbenzene and 1-vinylimidazole yielded a hydrophobic porous solid base, PDVB-xVI (x is the molar ratio of 1-vinylimidazole to divinylbenzene). The catalyst showed high catalytic activity in the methanol transesterification of both tripalmitin and ESG (Eruca Sativa Gars) oil with yields of 99.6 and 88.1%, respectively, which was in addition much higher than conventional base catalysts such as basic resin, hydrotalcite, CaO, and NaOH.[158] To enhance the basicity of catalysts, grafting an amine onto the skeleton of NIONs and then a following step of ion-exchange treatment of the as-obtained bromide-containing polymers in NaOH solution is an option.[104] Such solid base catalyst showed excellent performance in Knoevenagel condensation of benzaldehyde with ethyl cyanoacetate, together with high selectivity (99.9%), yield (99%), steady reusability (selectivity > 95%, yield > 95%), and resistance to water contamination.

### 3.1.2 NIONs with (organo)metallic group for catalysis

NIONs with an organic linker provide various opportunities for pre-synthetic or post-synthetic functionalization with metal ion as the active site for catalysis. The advantages of these chemically well-defined moieties within an open porous framework include uniform dispersion and good accessibility, which in fact can promote the catalytic activity in hydrogenation, oxidation, and so on. Such an advantage of a metal complex containing NION compared to the molecular derivative was proven by Lin *et al.*[24] In their work, NIONs with phosphorescent $[Ru(bpy)_3]^{2+}$ and $[Ir(ppy)_2(bpy)]^+$ building blocks were analysed. The porous polymer-immobilized complexes were efficient catalysts for light-driven reactions, such as the aza-Henry reaction (yield up to 99%), the R-arylation of bromomalonate (yield up to 91%), and the oxyamination of aldehydes (yield up to 48%), while the yields were comparable with, in some cases even higher than those of the homogeneous analogues (Fig. 17). Moreover, the heterogeneous catalysts were recyclable, underlining their robust nature. The same group also constructed porous polymerized tetraethynyl derivatives of $[Ru(bpy)_3]^{2+}$ *via* Eglinton homocoupling.[159] The obtained polymers with very high $[Ru(bpy)_3]^{2+}$ loading (up to 91.0 wt%) exhibited broad absorption bands between 300 and 800 nm together with relatively long excited state lifetimes (962 ns). These features made them valuable candidates as heterogeneous photocatalysts toward aza-Henry reactions, aerobic oxidative coupling of amines, and reductive dehalogenation reactions. A similar approach was used to incorporate Ir-, Rh-, and Re-bipyridyl complexes as well as cyclometalated Ir complexes into porous conjugated polymers for the investigation of reductive amination of ketones with high catalytic activities (yield up to 95%).[140]

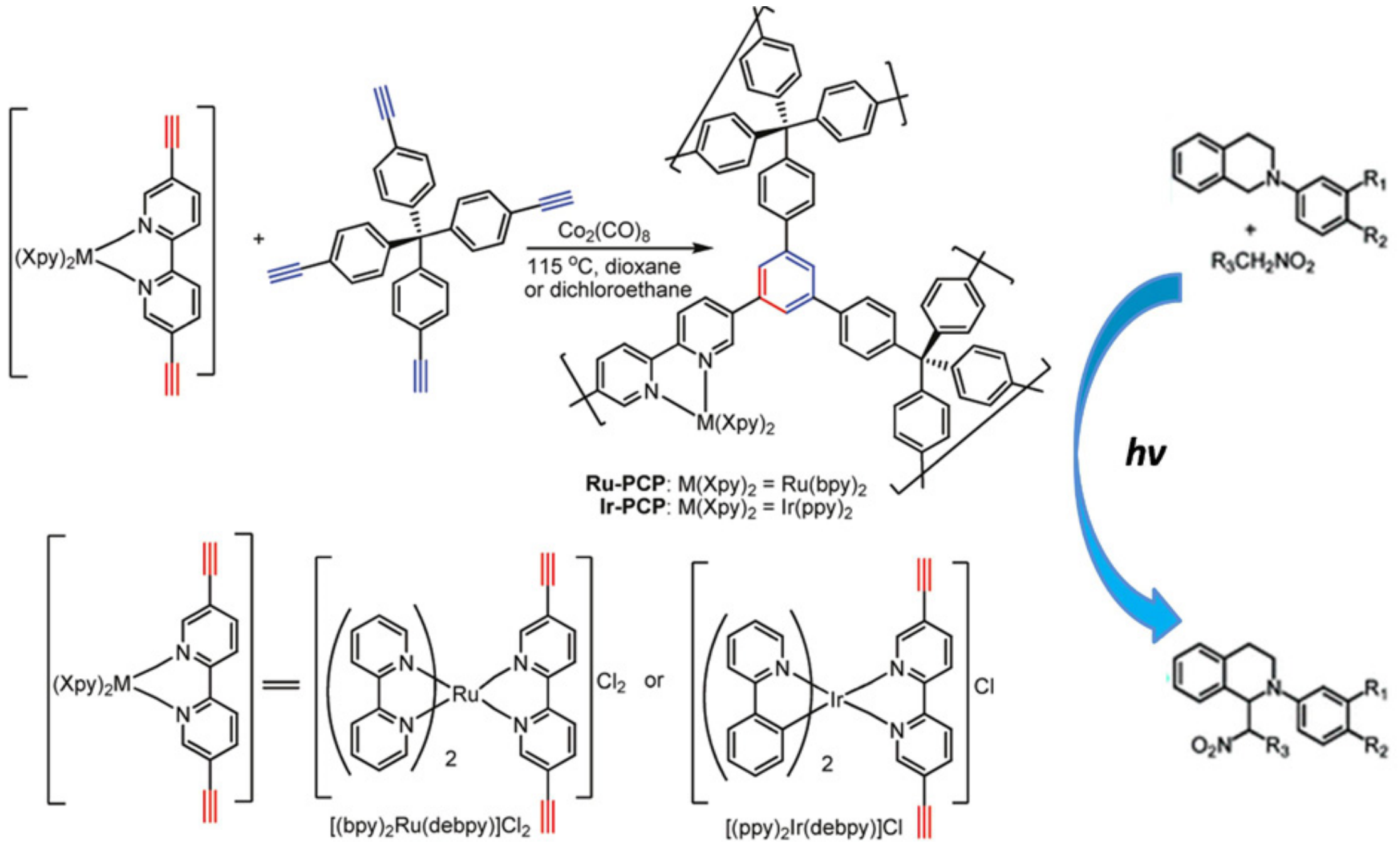

**Fig. 17** Synthesis of porous crosslinked polymers modified by charged units of $[Ru(bpy)_3]^{2+}$ and $[Ir(ppy)_2(bpy)]^{+}$ for photocatalytic applications. Reprinted with permission from ref. 24. Copyright 2011 American Chemical Society.

Recently, Li *et al.* reported a series of N-heterocyclic carbene-gold(I) functionalized porous organic polymers (Au-NHC@POPs) for heterogenous catalysis.[93] The materials were found to be efficient for alkyne hydration reactions with satisfactory tolerance towards catalysis environment and good recyclability at least for five times. Moreover, catalytic efficiency of Au-NHC@POPs with high porosity (yield: 86%) was much higher than a nonporous counterpart (yield: < 10%) of the same molecular composition. The combination of nanopores and homogeneously distributed, active metal-NHC sites obviously provides new opportunities in the exploration of built-in sustainable catalysts.

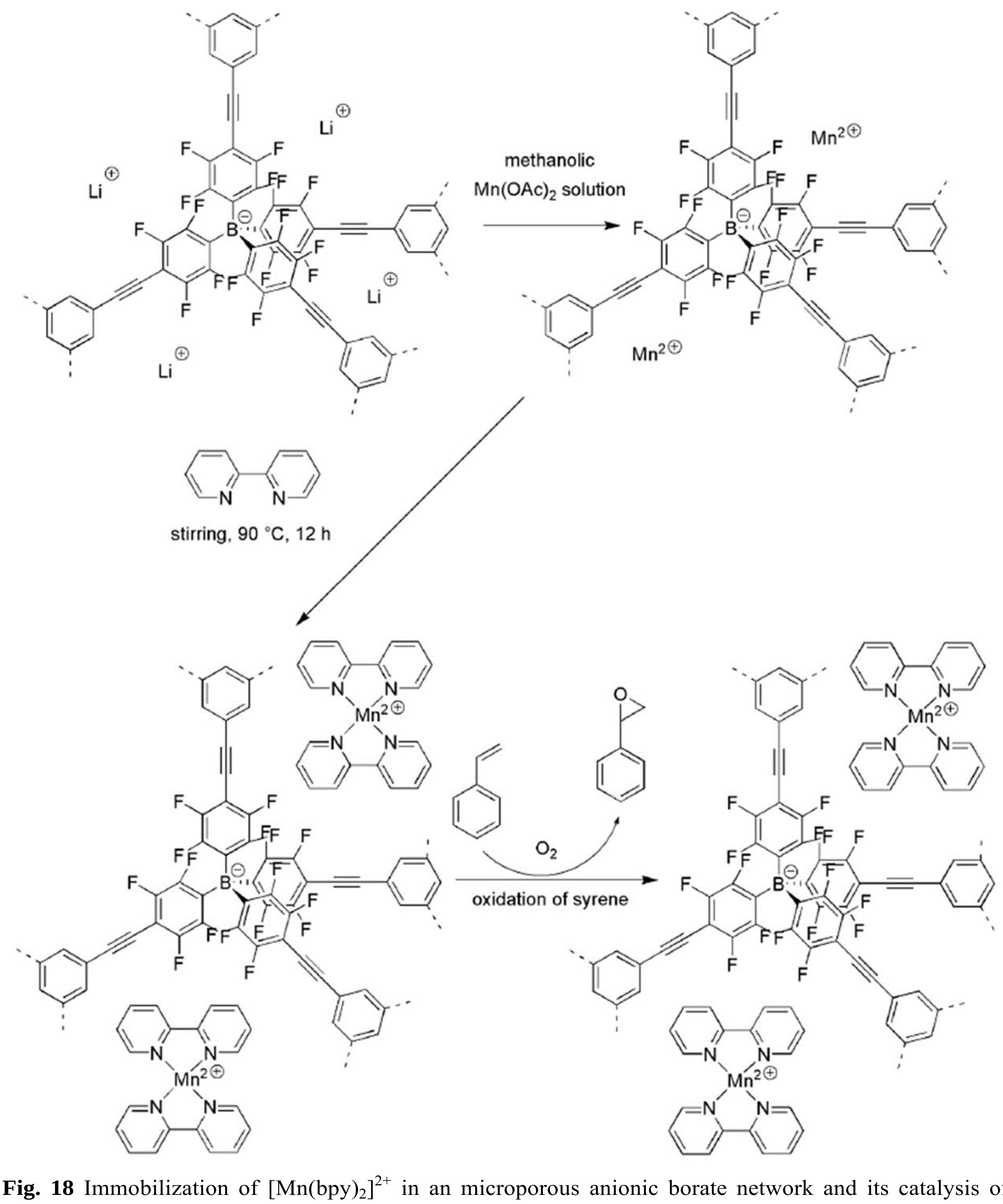

**Fig. 18** Immobilization of $[Mn(bpy)_2]^{2+}$ in an microporous anionic borate network and its catalysis of styrene oxidation. Reprinted with permission from ref. 40. Copyright 2013 Wiley-VCH.

Another prototypical family incorporating catalytically active metal sites for catalysis is provided by ion exchange of charged networks, *e.g.*, nanoporous polymers that consist of quaternary phosphonium ions as connection species between the aromatic linkers,[84, 160, 161] or negative borate ions.[40] In 2013, Thomas and co-workers developed an anionic microporous network containing weakly coordinating tetraphenylborate ions by copolymerization of $Li[B(C_6F_4Br)_4]$ and 1,3,5-triethynylbenzene *via* Sonogashira coupling.[40] The microporous architecture in combination with anionic skeleton was expected to make the metal counter cation mobile and fully accessible even in the solid state. As such, $[Mn(bpy)_2]^{2+}$ immobilized by NIONs, baptized as a "ship-in-the-bottle" approach, exhibited excellent catalytic activity toward aerobic oxidation of alkenes at 80 $^{o}$C in acetonitrile using dioxygen with selectivity for the formation of styrene oxide of 65 % and 100% conversion (Fig. 18). This approach posed great potential for the immobilization of catalytically active organometallic groups as long as suitable NIONs are available. Our group found that by simply refluxing micro-/mesoporous PIL powders with $CuCl_2$ in ethanol produced Cu(II)@PIL composite with an immobilized Cu(II) amount surprisingly being as high as 13 wt%.[36] The unexpectedly high capture ability of the host material towards $CuCl_2$ is based on multipole-induced ion pair adsorption of $CuCl_2$ onto the local imidazolium-based zwitterions in the PIL matrix. The material exhibited high performance in aerobic oxidation of hydrocarbons under comparably mild conditions with the yield and selectivity up to 98 and 99%, respectively. Such high catalytic activity was attributed to strong metal-support interaction generated by electrostatic bonding effect of ionic network, and the hierarchical micro-/mesopore supported mass transport during the catalytic process.

Normally speaking, NIONs with transition metal complexes could be possibly transferred to metal nanoclusters in the catalytic environment due to the metal leaching into solution. Therefore, the stability of transition metal complexes should be carefully investigated before the catalytic experiments. A method of quantitative determination of the coordinated metal species is necessary to specify the catalytic sites.

### 3.1.3 NIONs immobilized metal nanoparticles (MNPs) for catalysis

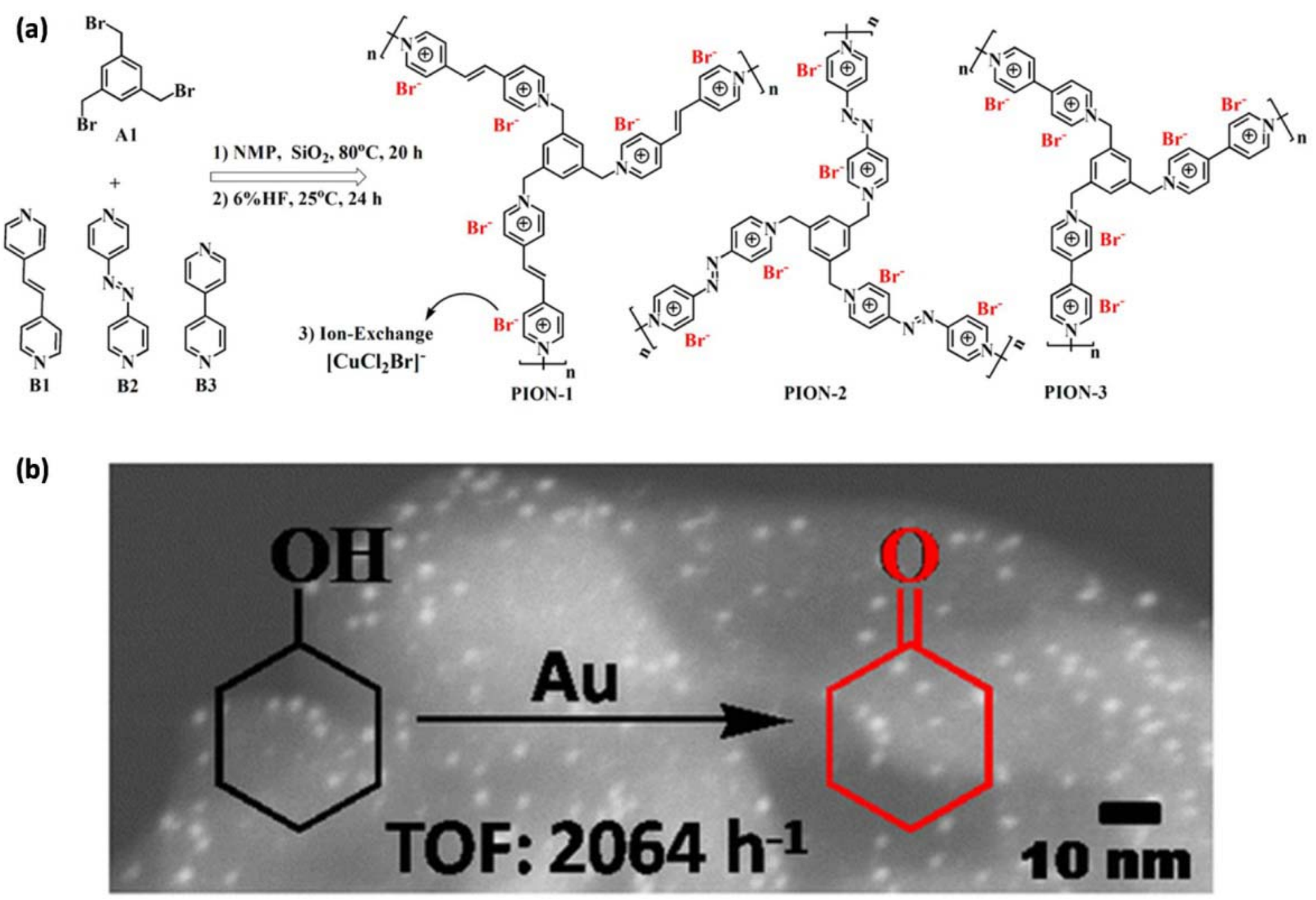

**Fig. 19** (a) Synthetic routes to NIONs, NMP: N-methyl-2-pyrrolidone. (b) STEM-HADDF image of an Au@NION sample for aerobic oxidation of saturated alcohols. Reprinted with permission from ref. 39. Copyright 2015 American Chemical Society.

NIONs due to their charged character could serve as both MNP stabilizers and support. In 2012, Zhang *et al.* firstly used ionic microporous network as support for immobilizing Pd nanoparticles.[84] The porous material was obtained by nickel(0)-catalyzed Yamamoto-type cross-coupling reaction with tetrakis(4-chlorophenyl)phosphonium bromide as the building unit. Due to the intrinsic ionic skeleton, tetrachloropalladate ($PdCl_4^{2-}$) precursor could be exchanged with bromide ion, followed by $NaBH_4$ reduction to obtain supported Pd nanoparticles. The catalyst exhibited excellent catalytic activity toward Suzuki cross-coupling reactions between chlorobenzene and phenylboronic acid (yield: 95.8%) as well as in the reaction between fluorobenzene and phenylboronic acid (yield: 87%). Besides the intrinsic nanoconfinement within the pores, highly polar ionic environment also allowed for the electrostatic stabilization of preformed nanoparticles, leading to a high activity of Pd nanoparticles. It is believed that the increase of the ion density could enhance the metal-support interaction, which could in turn boost the catalytic performance. Recently, Dai *et al.* developed a series of NIONs with extremely high ion density (three ion pairs per unit) based on polycondensation polymerization of a series of molecular knots (1,3,5-tris(bromomethyl)benzene (A1), 1,2-bis(4-pyridyl)ethylene (B1), 4,4′-azopyridine (B2), and 4,4′-bipyridine (B3)) in the presence of a hard template (silica particles ~12 nm in average size).[39] The $S_{BET}$ of resultant materials ranged from 107 to 132 $m^2$ $g^{-1}$ with hierarchical meso- and macropores, apart from the micropores (Fig. 19). This kind of porous materials exhibited strong ability to confine the Au nanoparticles in an ultrasmall size (1-2 nm) with homogeneous distribution due to the synergistic effect of nanoconfinement and electrostatic interaction with charged pores. The resultant materials exhibited superior catalytic activity and selectivity toward aerobic oxidation of saturated alcohol with a turn over frequency (TOF) of up to 2064 $h^{-1}$.

Besides the post-synthetic immobilization of MNPs onto NIONs, an alternative is the *in situ* generation strategy. For example, *in situ* generation of well-dispersed palladium nanoparticles of 2-3 nm in size immobilized by imidazolium-based NIONs could be obtained during the Suzuki – Miyaura cross-coupling polymerization process with $Pd(PPh_3)_4$ as the precursor.[162] The catalysts exhibited good activity (both the yield and selectivity are up to 100%) toward hydrogenation of nitrobenzene without extra addition of palladium species. The work opened a new avenue to the use of NIONs for MNP support and immobilization.

### 3.2 Environment related applications

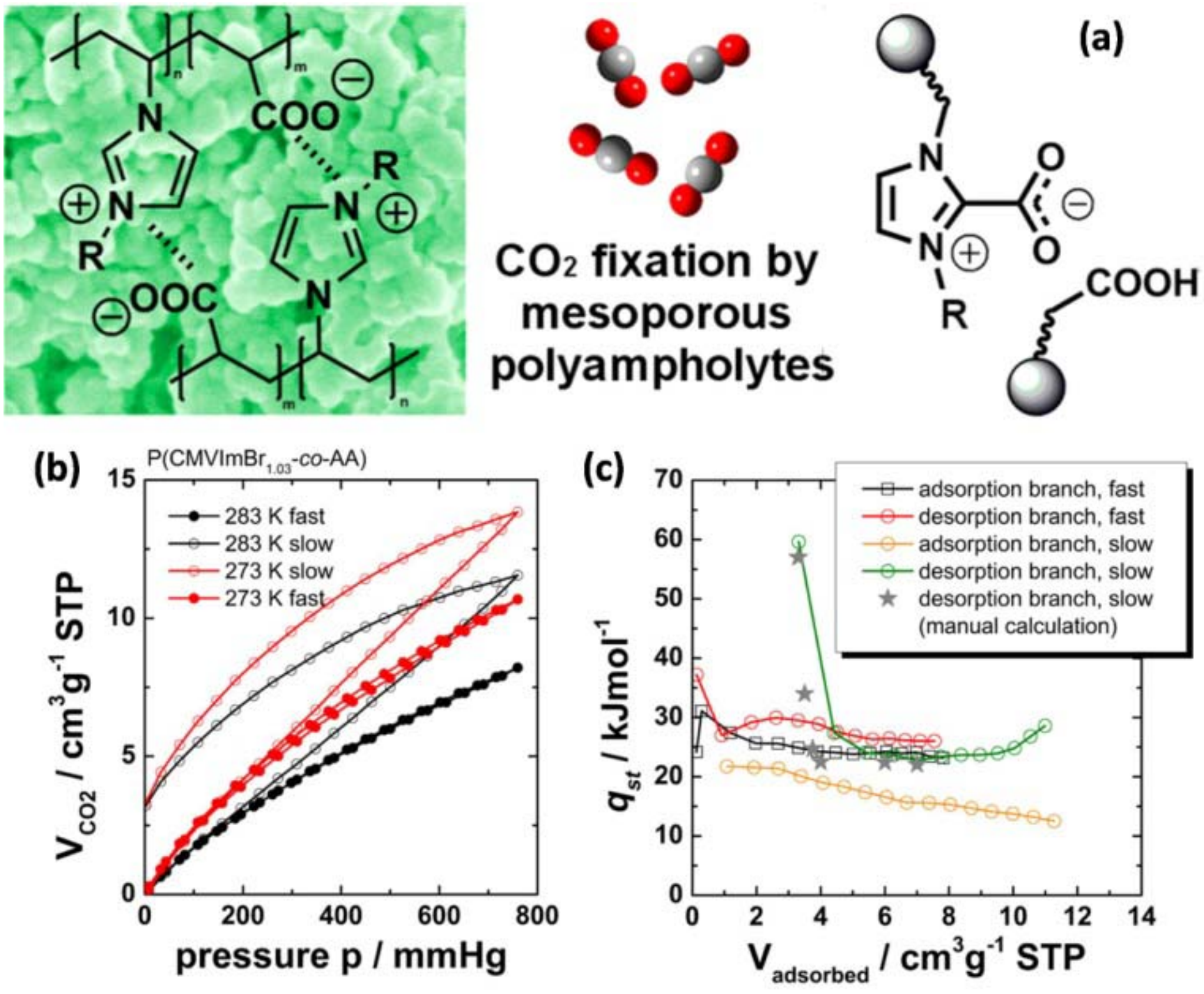

**Fig. 20** (a) Synthetic routes toward using mesoporous polyampholyte materials for $CO_2$ capture. (b) $CO_2$ adsorption/desorption isotherms of present porous material obtained at different temperatures and equilibration settings (closed symbols: fast; open symbols: slow); (c) isosteric heats of $CO_2$ sorption. Reprinted with permission from ref. 111. Copyright 2013 American Chemical Society.

Nanotechnology to tackle environmental issues plays a key role in enabling novel processes in environmental engineering and science, *e.g.* cost-effective technologies/materials for catalytic degradation, adsorptive removal and detection of contaminants.[163] As one of the key environmental issues of current research, a porous sorbent that can selectively capture $CO_2$ from a flue gas mixture to treat anthropogenic $CO_2$ emissions can be identified.[164-166] As compared to numerous kinds of porous materials that have been employed for $CO_2$ capture, NIONs with charge modified pore surface show a high selectivity toward $CO_2$.[167] Ionic pore walls can contribute to the build-up of Coulombic fields required for the polarization and polarized binding of polar molecules.[168] This is especially true when the pore size drops below 10 nm so that nanoconfinement effects strongly couple with the electrostatic interaction. Zhu and co-workers combined Materials Studio (MS) with grand canonical Monte Carlo simulation and revealed that the values of binding energy followed an order of $H_2$, $O_2$, $N_2$, $CH_4$ and $CO_2$ (small to large) on a series of quaternary pyridinium-type porous aromatic frameworks with tuneable channels.[37] This result is consistent with the observation in gas sorption experiment. The high value of $CO_2$ was ascribed to the distribution of a partial positive charge on the pyridinium groups, leading to a polarizing binding environment which enhances the affinity towards $CO_2$ owing to dipole–quadrupole interactions.[169] Besides the physical absorption process, $CO_2$ could also be absorbed through chemisorption processes in IL-based materials due to their strong interaction and the preferred formation of imidazolium-carboxylates, formally *via* a transient N-heterocyclic carbene intermediate. Especially materials with small pores and imidazolium species are predicted to be highly efficient for $CO_2$ capture. An interesting example of the coupling of physical and chemical absorption of $CO_2$ was studied in a recent report on mesoporous imidazolium-type PIL-based polyampholytes (Fig. 20).[111] These materials exhibited an intrinsic porosity with $S_{BET}$ of up to 260 $m^2\ g^{-1}$. It was found that next to fast $CO_2$ adsorption to the surface of ionic network (which is the predominant mechanism in typical micro-/mesoporous organic polymers), additional volume uptake and swelling deep into the polymeric matrix occurs. This process is slow compared to the surface adsorption and comes with an energetic penalty. Moreover, $CO_2$ taken up as such could not be desorbed easily. Fourier transform infrared spectroscopy (FTIR) measurement provided evidence that the trapped $CO_2$ could partially be activated to form imidazolium-carboxylate zwitterions even at low temperature and $CO_2$ pressure. The mechanism was believed to be comparable with the previously reported formation of transient N-heterocyclic carbene intermediate within low-molecular ILs of the imidazolium-carboxylic anion type.

The NIONs with a more delicately modified pore surface could also be used for removal of ammonia gas from industrial air effluents which are basic in nature. For example, a NION was prepared by using PAF-1[20] with post-grafted –COOH groups, and it was tested for the selective removal of ammonia.[170] The material featuring a multiple interpenetrated structure dominated by micropores (< 6Å), exhibited an uptake of 17.7 mmol $g^{-1}$ at 1 bar, which was the highest capacity in this application reported so far. Such exceptional performance could be attributed to multiple chemical interactions between the multiple acidic sites located in the host material and ammonia gas, which promoted the adsorption for pollutant capture.

Ionic exchange materials for adsorption of ionic pollutants or heavy metals are another class of relevant sorption materials, dominated by conventional ion-exchange resins.[171] Conventional ion exchange resins with large sized pores (in the range of several to tens of micrometers) face a number of drawbacks, such as inefficient accessibility of ion-exchange sites, limited kinetics and sometimes "outflow" of mobile phase under working conditions.[171] The NIONs with rigid skeleton as well as intrinsic nanosized pores already have shown applicability toward environmental pollution such as radioactive ions,[86] heavy metal ions,[95, 172] organic dyes,[110] and so on. In this regard, Dai and co-workers showed that NIONs primarily operate by an ion exchange mode as demonstrated by imidazolium containing NIONs to capture perrhenate ions ($ReO_4^-$).[86] Energy-dispersive X-ray analysis (EDAX) results showed that in contact with a $NaReO_4$ solution, the porous network did not contain any $Na^+$, reemphasizing that perrhenate is taken up through an ion-exchange process involving only the counteranions. Sometimes the micropore structure of the ionic networks also enhances the ability in ion capture. For example, high surface area of PAF-1 functionalized with a thiol group could work as a nano-trap for mercury and showed a record uptake capacity of mercury over 1000 mg $g^{-1}$.[172] Such extremely high affinity towards Hg(II) stemmed from the strongly chelating deprotonated framework of PAF-1-$S^-$ coupled to the rigid scaffold with high accessible surface area. A similar strategy by grafting carboxylate groups on PAF generated a material with high metal loading capacities toward $Sr^{2+}$, $Fe^{3+}$, $Nd^{3+}$, and $Am^{3+}$, from aqueous solutions together with excellent adsorption selectivity for $Nd^{3+}$ over $Sr^{2+}$, as required in the treatment of radioactive waste.[95]

A convinced case of using charge-modified porous networks for radiological iodine capture was demonstrated recently. A series of NIONs (PAF-23, PAF-24, and PAF-25) were built up from a tetrahedral building unit, lithium tetrakis(4-iodophenyl)borate (LTIPB), and different alkyne monomers as linkers *via* a Sonogashira–Hagihara coupling reaction (Fig. 21).[83] The networks featured three effective sorption sites, *i.e.* an ionic site, the phenyl ring, and triple bonds, and exhibited the highest reported iodine adsorption capability to date (2.71 g/g, 2.76 g/g, and 2.60 g/g of iodine for PAF-23, PAF-24, and PAF-25, respectively.). A control experiment was carried out by using neutral PAFs with similar topology for $I_2$ capture under the same conditions, which ended up with a much lower capacity.

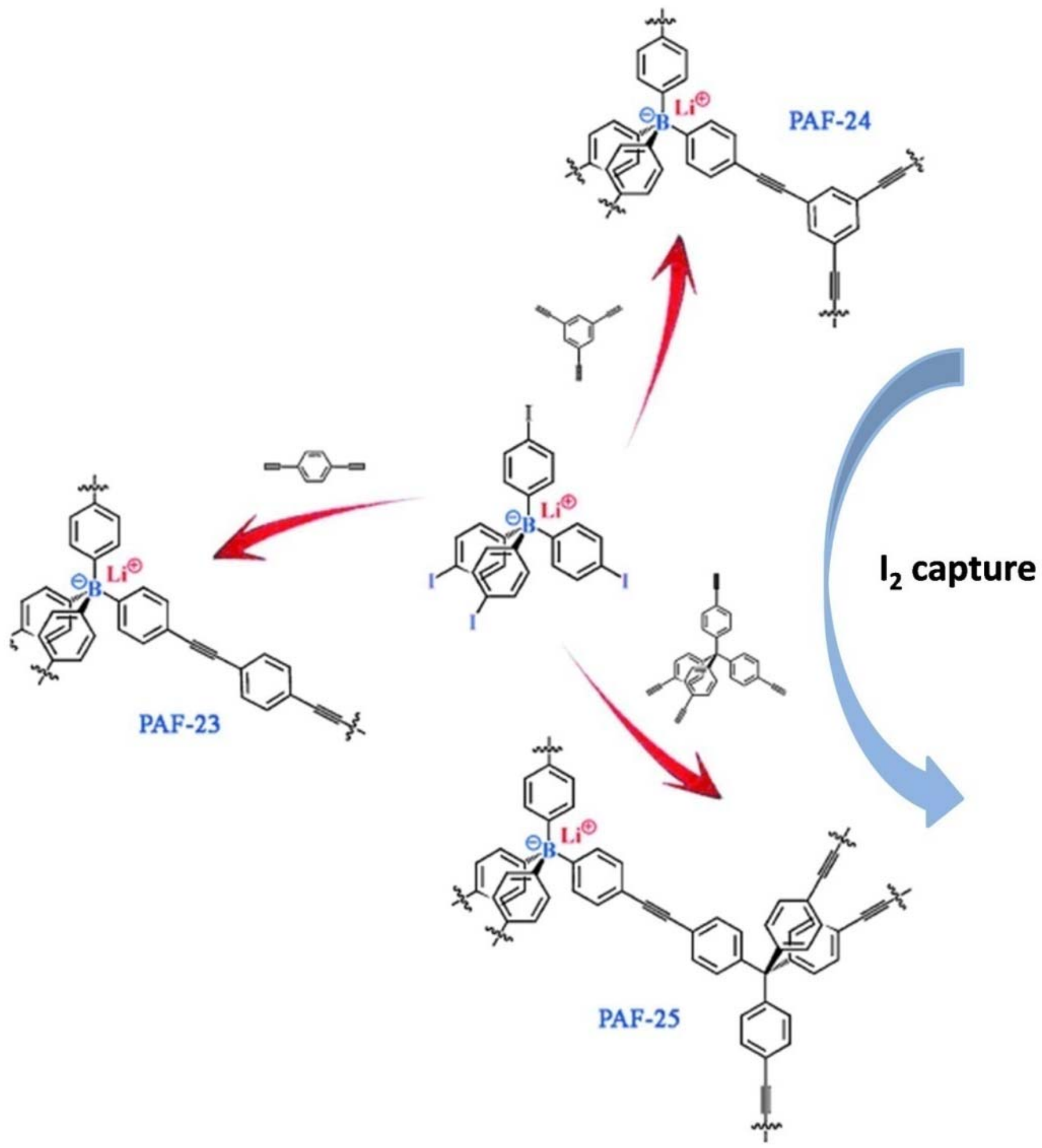

**Fig. 21** (a) Synthetic routes for ionic porous polymers PAF-23, PAF-24, and PAF-25 by Sonogashira–Hagihara coupling reactions and their use for $I_2$ capture. Reprinted with permission from ref. 83. Copyright 2015 Wiley-VCH.

Membrane-based technologies for the production of pure water from seawater or brackish water are an energy-efficient, low environmental impact engineering solution. NION-membranes are potentially a promising material for such application.[173,174] Zhang and co-workers reported a nanoporous membrane (pore size ranging from tens to hundreds of nanometers) through ionic complex of poly(acrylic acid-co-acrylonitrile)s and imidazolium-based polycations. Zeta potential measurements indicated that the obtained membranes were negatively charged at neutral pH conditions. The membranes with hierarchically structured nanopores exhibited moderate rejection to salts in the order of $Na_2SO_4$ (59.5 %) > NaCl (6 %) > $MgCl_2$ (0.1 %), but high rejection to methyl orange (> 99.9 %).[173]

Organic solvents are inevitably utilized in chemical industry, whilst quite some of them are not expected to show up near the consumer. Although various methods have been explored for solvent sensing, the exploration of smart actuating materials that are capable of adaptive motion, and/or reversible shape variation in response to solvent stimuli is an exotic option.[175,176] Currently, most reported polymer actuators toward solvent sensing applications suffered from a relatively low sensitivity for organic solvents due to the requirement of a substantial amount of secondary solvents to produce noticeable shape deformation or displacement. Our group recently reported a nanoporous (30–100 nm in pore size) PIL membrane prepared by ionic complexation between $PCMVImTf_2N$ and PAA, which carried a unique gradient of crosslinking density along the membrane corss-section.[116] The membrane in water readily bent upon adding as low as 0.25 mol% of acetone molecules (1 acetone per 400 water molecules). This is at least one order of magnitude more sensitive than other state-of-the-art solvent stimulus polymer actuators (SSPAs). Mechanism investigation revealed that the strong interaction between acetone and the ion pair in $PCMVImTf_2N$ together with the coexistent structural gradient led to a gradient absorption of acetone along the cross-section of membrane, resulting in a swelling gradient across the membrane to bend the membrane. Moreover, different from common nonporous SSPAs, the nanoporous channel not only accelerated mass transport of solvents into the membrane, but also weakened the overall bending rigidity by introducing a gas subphase.

## 3.3 Energy storage and conversion

As a new type of porous polymer, NIONs have been explored for various energy-related fields, *e.g.*, carriers for enhanced gas storage, electrode materials for batteries, supercapacitors, and fuel cell membrane, over the past few years. In this section, we intend to catch this emerging field, illustrating the unique role of charge containing units combined with pores of different sizes on the performance of these materials.

### 3.3.1 NIONs for gaseous fuel storage

One of approaches for the energy storage of porous materials is the direct deposition of energy molecules ($H_2$, or $CH_4$ gas) inside the pores. A material with tiny pores (micropores and small mesopores) is prerequisite for this application, as then capillary pressure is higher, *i.e.* more gas can be stored while not changing any storage parameters. Although there are numerous porous materials reported for gas storage, NIONs show unique opportunities for this application. Worth to note, theoretical calculation proposed that metal ions (*e.g.*, $Li^+$, $Mg^{2+}$) containing porous frameworks would enhance $H_2$ adsorption energy.[177] In fact, one major advantage of metallated NIONs is that the metal site on the pore wall/skeleton are highly accessible, which is not easily achieved by state-of-the-art $H_2$ storage MOFs in which the metal site are shielded by organic linkers.

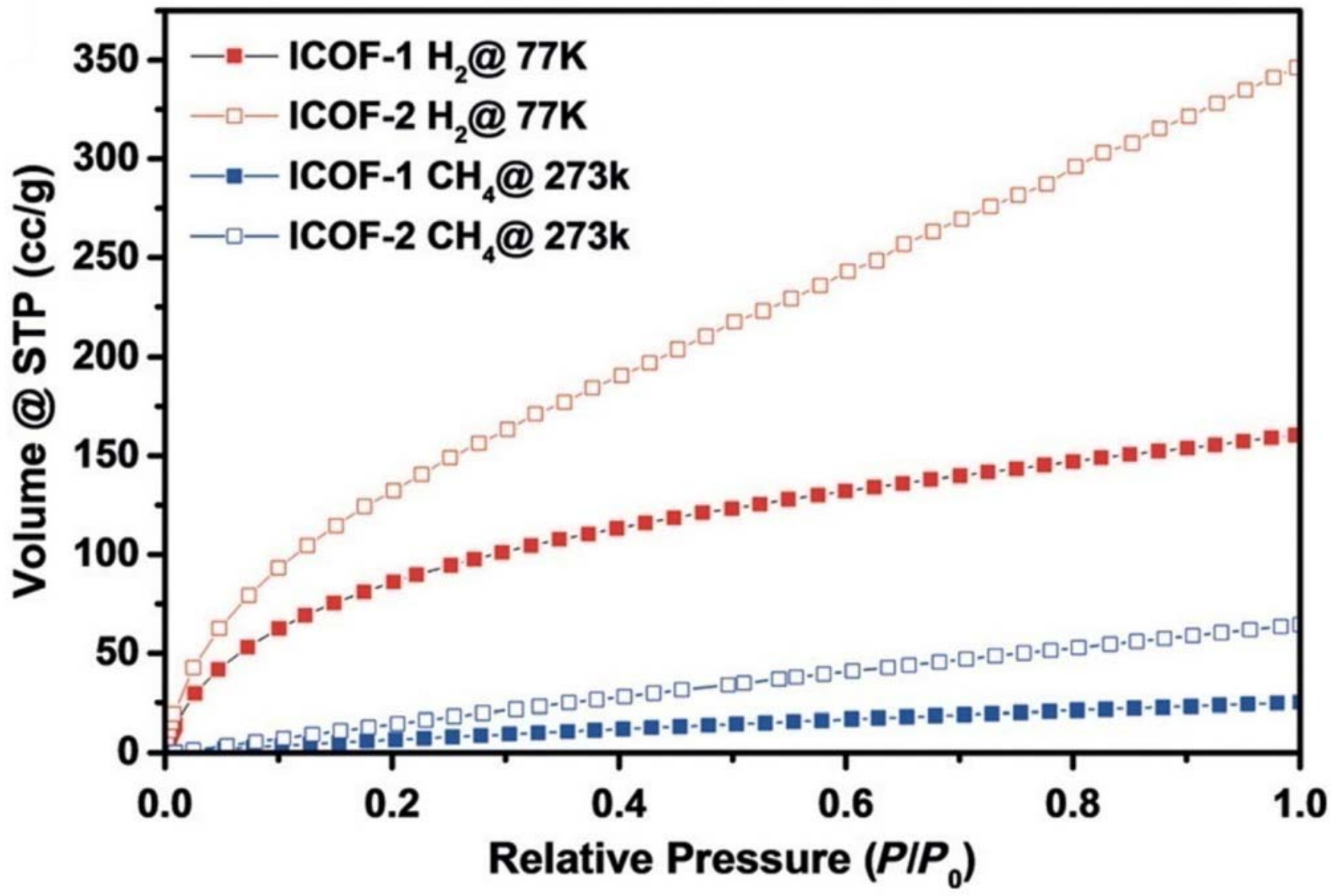

**Fig. 22** Hydrogen and methane adsorption isotherms for ICOF-1 and ICOF-2. Filled symbols represent adsorption and hollow symbols represent desorption. Reprinted with permission from ref. 50. Copyright 2016 Wiley-VCH.

Cooper *et al.* found that tungsten-based organometallic complex, which is known to form strong bonds to hydrogen, could be incorporated into the polymer support for hydrogen generation.[178] Interestingly, the storage and release of hydrogen could be triggered by UV light at modestly cryogenic temperatures (220 K) or above. The work indeed suggested that metal complex containing NIONs is promising for enhanced hydrogen storage even at ambient temperature due to the metal–$H_2$ bond energies that may fall in the desirable range for reversible $H_2$ storage at close to ambient temperatures. Ma *et al.* used carboxyl-decorated PAF-26-COOH as precursor to prepare NIONs *via* a post-metalation step using $Li^+$, $Na^+$ and $K^+$ ions.[136] These materials featuring micropores and moderate surface areas (ranging from 430 to 572 $m^2$ $g^{-1}$) exhibited increasing uptake of $CH_4$ compared with the unmodified porous network from 35 (34 $cm^3$ mg) to 76 wt% (60 $cm^3$ mg). This increase in $CH_4$ uptake followed the enhanced initial $Q_{st}$ values (from 14.3 to 24 kJ $mol^{-1}$) of $CH_4$ in comparison with the noncharged PAF to the ionic PAFs. Such strong affinity towards $CH_4$ was attributed to charged pores that facilitated polarization of the $CH_4$ molecule and resulted in stronger binding force. Similar trends have been observed as well in ion-functionalized nanoporous membrane system.[137] An enhanced performance could be realized both by increasing $S_{BET}$ as well as ionic density in the polymer network. For example, an ionic covalent organic framework (ICOF-2), which contained $sp^3$ hybridized boron anionic centers and tuneable countercations, was constructed by formation of spiroborate linkages (Fig. 22).[50] This material featuring high surface area (1259 $m^2$ $g^{-1}$) showed exceptional large adsorption capacities towards $H_2$ (3.11 wt%, 77 K, 1 bar) and $CH_4$ (4.62 wt%, 273 K, 1 bar), among the highest of all organic porous materials that have been so far analyzed. By contrast, the ICOF-1 with $[Me_2NH_2]^+$ cation showed low performance toward gas capture ($H_2$: 1.42 wt%, 77 K, 1 bar; $CH_4$: 1.65 wt%, 273 K, 1 bar), illustrating that the superior properties of ICOF-2 indeed can be attributed to the high concentration of $Li^+$ counterions in the pore, which can significantly increase the gas uptake by increasing the molecular binding energy.

### 3.3.2 NIONs for photoelectrochemical energy storage and conversion

Porous organic networks decorated with charge units will lead to high mobility or biased transport of the electron/hole in the skeleton, especially when $\pi$-conjugated networks are considered. This kind of materials are photo/electrochemically active,[50-52, 90] and show intriguing potential in certain energy related applications.

A tutorial and conceptual case of this kind of materials is ionic COFs with a highly ordered 2D open network structure, which has the capability to predictably organize redox-active groups. This fact makes them potential candidates for charge storage devices. For example, Dichtel and co-workers reported a β-ketonamine-linked 2D COF containing redox-active anthraquinone building unit.[179] The material exhibited a moderate capacitance of 48 ± 10 F $g^{-1}$ at a current density of 0.1 A $g^{-1}$ (1 M $H_2SO_4$), which after 5000 cycles only dropped to 40 ± 9 F $g^{-1}$. It should be mentioned that the chemical and oxidative stability of COF linkages are mandatory for the direct use in electrochemical devices, and thus a careful design is advised. Recently, Jiang and co-workers reported a facile and general strategy that converts a conventional COF into redox-active platform by delicate post-synthetic channel-wall functionalization with organic radicals (Fig. 24).[180] In their work, the conventional imine-linked COF ($[HC{\equiv}C]_{X\%}$-NiP-COF; X = 0, 50, and 100) as a scaffold with nickel porphyrin at the vertices and ethynyl unit on the channel wall was synthesized.

The ethynyl groups of this structure were then "clicked" to 4-azido-2,2,6,6-tetramethyl-1-piperidinyloxy in a smooth and clean manner to yield $[TEMPO]_{50\%}$-NiP-COF and $[TEMPO]_{100\%}$-NiP-COF quantitatively. Organic TEMPO radicals do not only hold all the unique properties of radicals but also feature reversible switching between the oxidation states of the neutral radical and the oxoammonium cation. Cyclic voltammetry (CV) measurements demonstrated the redox-active nature of these ionic COFs. The galvanostatic charge and discharge tests revealed that the $[TEMPO]_{100\%}$-NiP-COF exhibited a capacitance of 167 F $g^{-1}$ at 100 mA $g^{-1}$.

Polymer electrolytes are a key component for solid electrolyte batteries. The popular solid electrolyte candidates as complexes of Li salts and polyethylene oxide generally have low conductivities at room temperature ($<10^{-5}$ S $cm^{-1}$).[181,182] It is well-known that the ionic conductivities in solid electrolytes were governed by both the segmental motion of the chains and the number of dissociated carrier ions and their mobility.[183] NIONs with high density of ion pairs as well as pore accessibility are suitable candidates for ionic conductors. Dai and co-worker reported an intriguing PIL-based ionic network by a direct nucleophilic substitution reaction between hexakis(bromomethyl)benzene and 4,4′-bipyridine.[184] This polymer featured high charge density (six ion pairs per repeating unit) with good ionic conductivities (up to $5.32 \times 10^{-3}$ S $cm^{-1}$ at 22 °C). Moreover, such network based solid electrode exhibited wide electrochemical stability windows up to 5.6 V, and good interfacial compatibility with the electrodes. An initial high discharge capacity up to 146 mA h $g^{-1}$ at 25 °C was found in Li/$LiFePO_4$ battery assembled with such ionic network-based electrolyte.

NION-based materials with enhanced ion conduction could also be important in fuel cell applications.[131,185] For example, mesoporous poly(benzimidazole) (PBI) membranes with pore sizes of ~10 nm were obtained by hard templating method.[131] The post-grafted phosphoric acid groups to yield a highly proton conducting material at zero humidity could be easily operated up to 180 $^{o}$C. Moreover, the proton conductivity was one to two orders of magnitude higher than that of a non-porous PBI/$H_3PO_4$ complex, as tested under similar conditions. Such excellent performance was related to the nanopores of PBI/$H_3PO_4$ membrane. The interconnected pores provide the proton conduction highways, and meanwhile excessive membrane swelling was prohibited by a high crosslinking density of the membrane, thus the membrane was also mechanically stable. Another example of enhanced conductivity in a charged porous polymer network was given by using graphitic carbon nitride ($g$-$C_3N_4$)[27] as a polymer precursor through post protonation with HCl. The resultant protonated material ($g$-$C_3N_4$-$H^+Cl^-$) showed at least 10 times enhancement in ionic conductivity as compared with parent $g$-$C_3N_4$ (Fig. 23).[186]

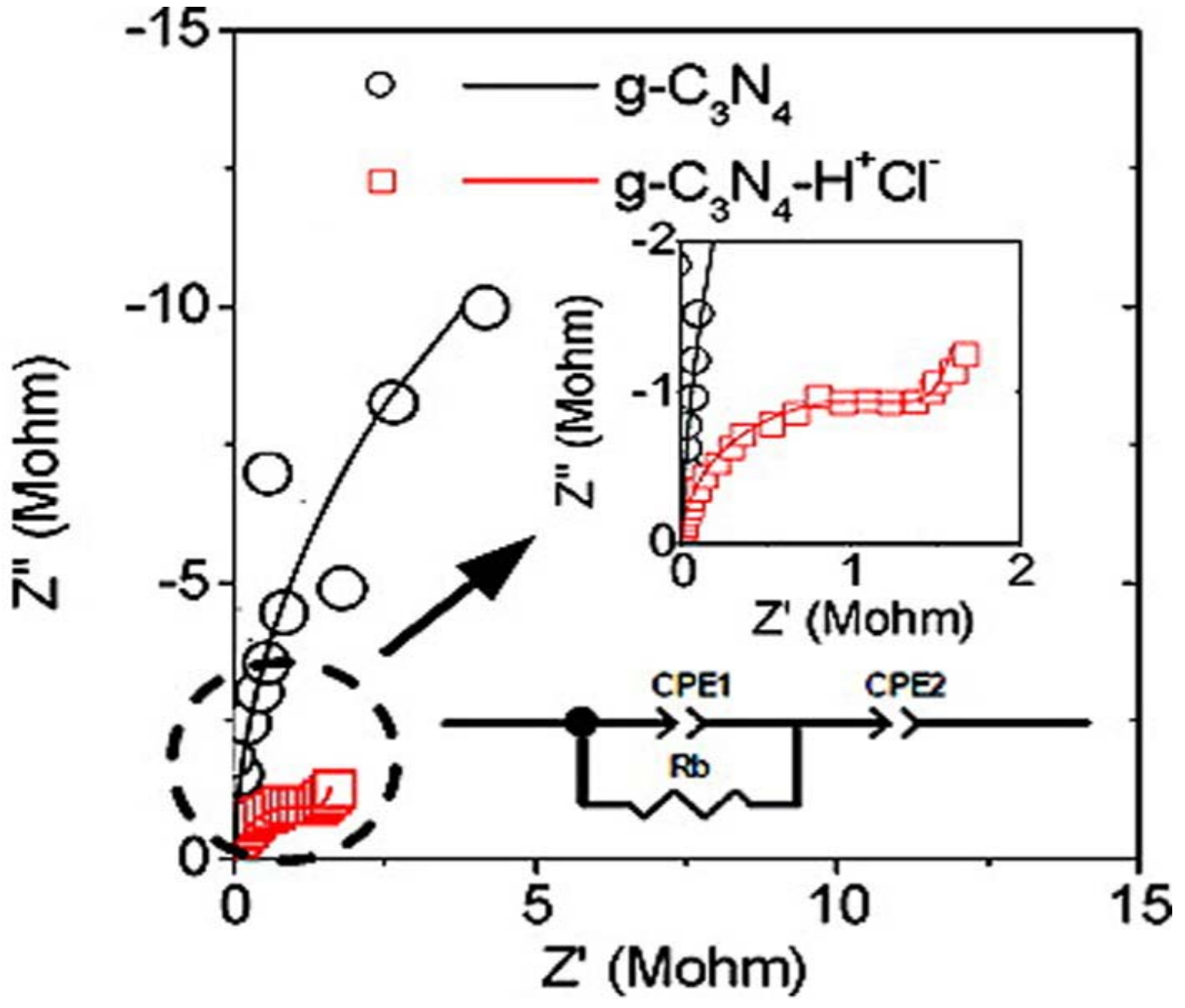

**Fig 23**. Nyquist impedance plots (scatters) for g-$C_3N_4$ and g-$C_3N_4$-$H^+Cl^-$ and simulation (lines). The frequency range is from 106 to 103 Hz, and the perturbation signal is 100 mV. Inset: dotted area in high magnification and equivalent circuit mold. The calculated resistances ($R_b$) before and after protonation were ca. 28 and 1.5 MΩ. Reprinted with permission from ref. 186, with the Copyright 2009 American Chemical Society.

Integration of redox process in the NION was already discussed to be relevant to energy storage applications. Dichtel *et al.* found that when naphthalene diimides (NDIs) were incorporated into microporous polymer films, the carbonyl groups of adjacent NDI radical anions and dianions bound strongly to $K^+$, $Li^+$, and $Mg^{2+}$, consequently shifting the formal potentials of NDI's second reduction by 120, 460 and 710 mV for $K^+$, $Li^+$ and $Mg^{2+}$-based electrolytes, respectively. By contrast, the formal reduction potentials of NDI derivatives in solution state didn't show any difference toward $K^+$-, $Li^+$- and $Mg^{2+}$-based electrolytes, and such shifts have not been reported for NDI previously.[187] These results implies the profound influence of the solid-state structure of a NION on its electrochemical response. The microstructure control is significant in general in electrical energy storage, and one can envision that those shifts allow for generation of devices operating at higher voltages to increase energy density.

Conjugated NIONs could be explored for their optoelectronic effects. Marken, McKeown and coworkers found that microporous polyamine prepared by a polymerization reaction involving the formation of Tröger′s base possessed the capability to switch from protonated anion-conducting to neutral anion/cation-conducting behavior.[188] The suppressed proton conductivity in the neutral state was speculatively attributed to isolated amine sites and exceptional chain rigidity of the polymer, which was responsible for the generation of an ionic diode.

Electrode interlayer is a key structure between active layers and conducting electrodes that controls the transport of charge carriers in and out of cells. Jiang and co-workers employed a microporous network bearing polyborane carbazole unit as a new type of electrode interlayer.[189] They have found that the neutral network based thin-film exhibited extremely low work-function-selective electron flow; while upon ionic ligation and electro-oxidation, the charged network significantly increased the work function and turned into a hole conductor. Moreover, these charged thin films were compatible with various electrodes and offered outstanding functions in various types of devices, including solar cells and light-emitting diodes.

Salinity difference between sea water and river water creates an exploitable salinity gradient energy, so-called “blue energy”, a clean energy source popularly discussed in the current energy crisis.[190] In principle, more than two terawatts of electricity can be potentially generated in the river estuary where the rivers flow into the sea. To capture this energy more efficiently, numerous efforts have been made to create mew materials enabling this transformation. Porous materials with high surface charge density and pore-sizes down to sub-10 nm are favorably discussed in this context.[190,191] Recently, by integrating a porous block copolymer membrane $PS_{48400}$-*b*-$P4VP_{21300}$ supported by a porous polyethylene terephthalate (PET) substrate, Jiang *et al.* developed a nanofluidic generator with ultrahigh rectification ratio (~ 1,075). By flowing an aqueous solution of 0.5 M NaCl into 0.01 M NaCl through the porous membrane, the heterogeneous nanoporous ionic membrane is capable of outputing a maximum power density of 0.35 $W \cdot m^{-2}$, exceeding some commercially available cation exchange membrane.[192]

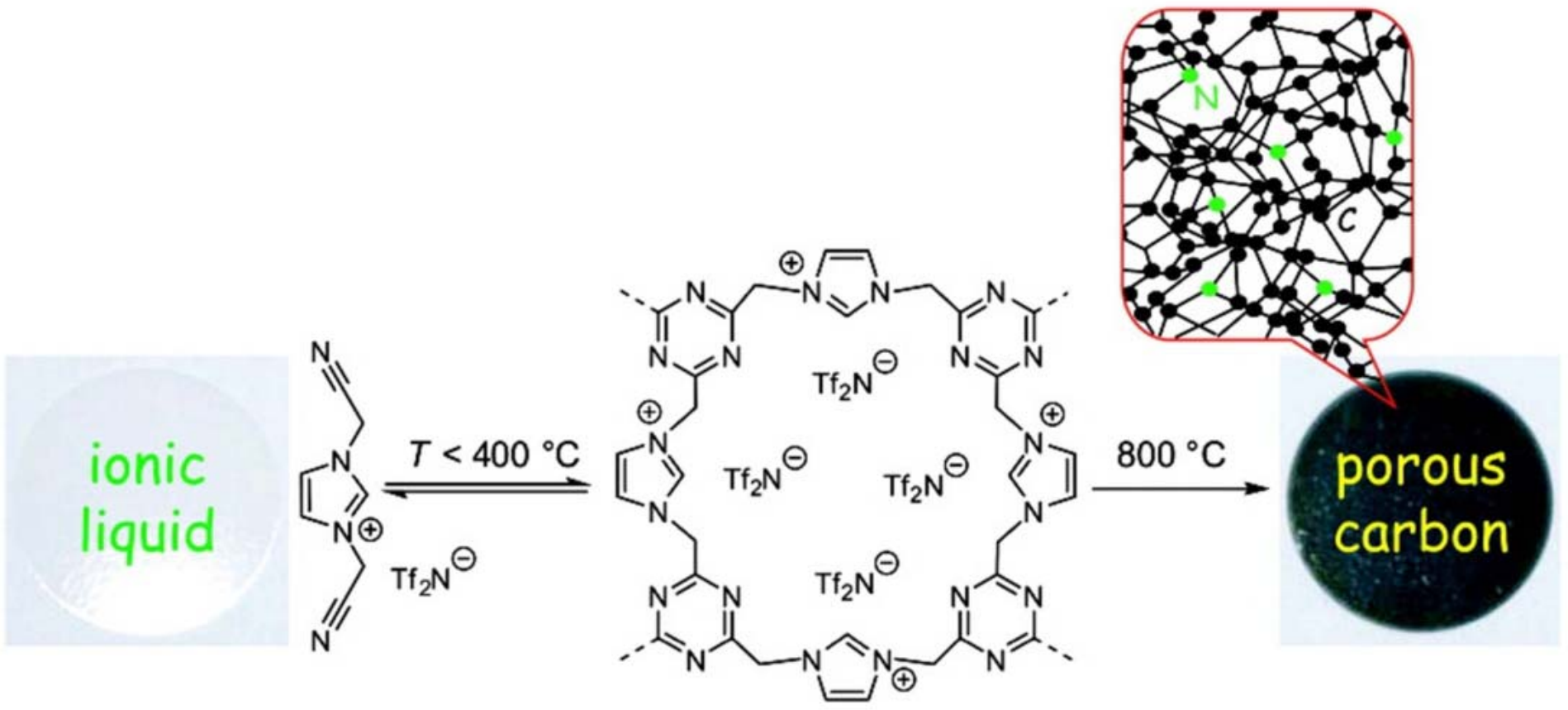

**Fig. 24** Schematic routines toward the synthesis of porous carbons through polymerization of IL precursors. Reprinted with permission from ref. 195, with the Copyright 2009 American Chemical Society.

Besides the direct use of NIONs for energy storage and conversion, NIONs could be used as precursors for the preparation of carbonaceous materials, which are ubiquitous in various technological and energy-related applications. Carbon-rich NIONs are intrinsically suited precursors for preparing carbon nanomaterials. Compared with conventional polymeric precursor, charged polymers with intrinsic ionic nature and strong Coulombic interactions in ionic pairs lead to high thermal stability of precursor, which reduces mass loss before the decomposition process begins.[193,194] Dai *et al.* investigated the porous carbons through polymerization of IL precursors (Fig. 24).[195] The nitrile functionalized cations 1,3-bis(cyanomethyl)imidazolium ($[BCNIm]^+$) underwent a cyclotrimerization and formed a dynamic amorphous porous polytriazine networks with temperature around 400 ℃. The resulting textural properties such as pore structure and surface area were also tuneable based on the structural motifs of the ions. These important findings indicated a new avenue for producing carbon materials from NION templates and precursors under a maximal possible retention of a previously engineered shape and texture.[196]

The advantage of using NIONs as carbon precursors not only lies in the abundant carbon content in the skeleton, but also the ionic components, which could serve as guest species, yielding through carbonization multifunctional porous carbons or carbon hybrids. Han *et al.* reported a cationic, phosphorous-based NION as a template and precursor for carbonization (Fig. 25).[197] Before the calcination, transition-metal-containing anions, such as tetrathiomolybdate ($MoS_4^{2-}$) and hexacyanoferrate ($Fe(CN)_6^{3-}$), were loaded into the porous network to replace the original iodide anions, resulting in polymer networks containing complex anions (termed HT-Met, where Met = Mo or Fe). After pyrolysis under a hydrogen atmosphere, the HT-Met materials were efficiently converted at a large scale to metal-phosphide containing porous carbons (denoted as MetP@PC). These composites exhibited superior electrocatalytic activity for the hydrogen evolution reaction (HER) under acidic conditions. Particularly, MoP@PC (0.24 mg $cm^{-2}$ on a glass carbon electrode) exhibited an overpotential of 51 mV at 10 mA $cm^{-2}$ and a Tafel slope of 45 mV $dec^{-1}$, comparable to those of the commercially available Pt/C catalyst (24 mV at 10 mA $cm^{-2}$; 30 mV $dec^{-1}$).

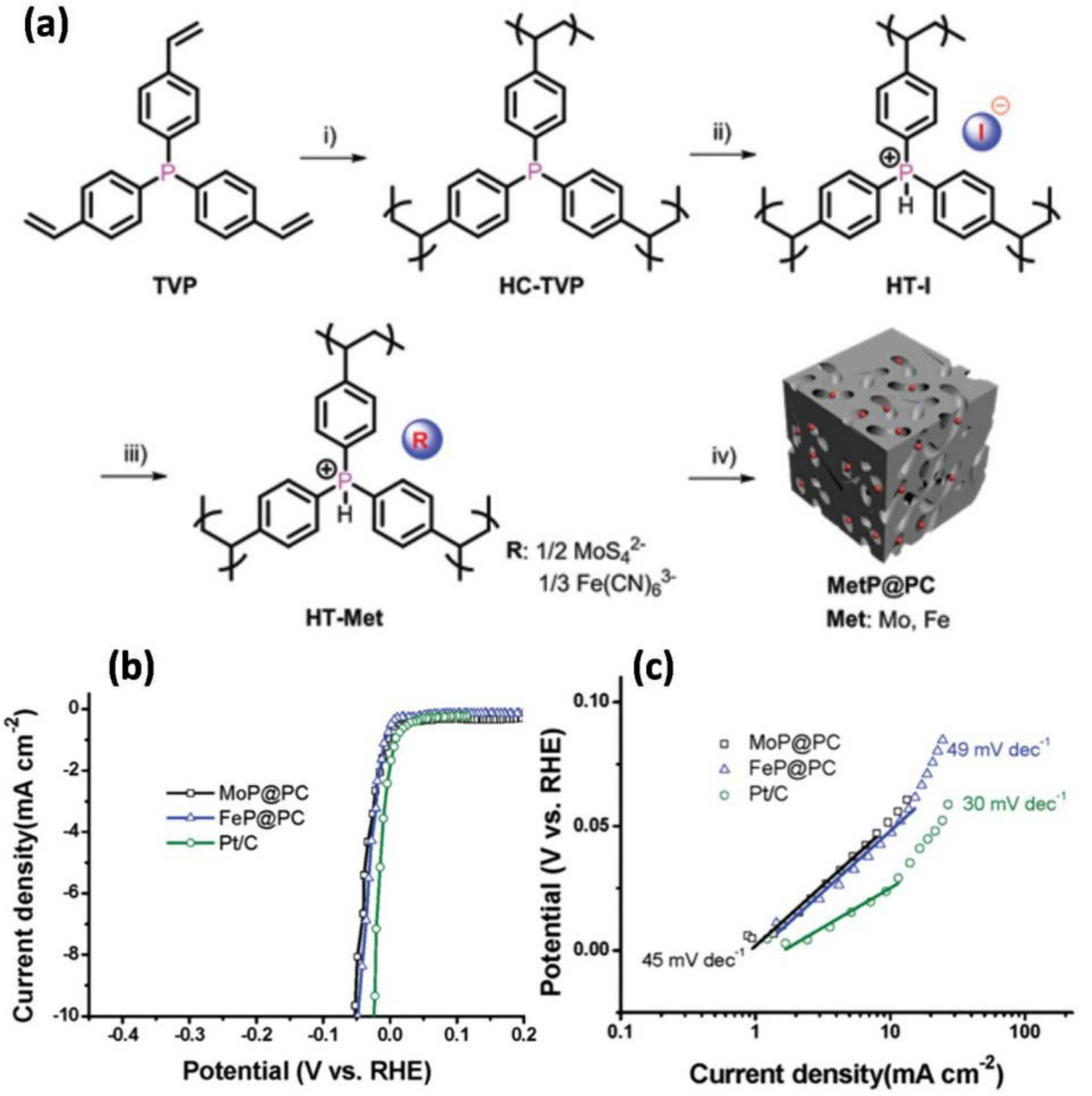

**Fig. 25** (a) Preparation of NION and related mesoporous carbons embedded with metal phosphides (MetP@PCs). Reaction conditions: i) AIBN, tolune, 65℃, 24h; ii) dioxane, aqueous hydroiodic acid, room temperature, 24 h; iii) $(NH_4)_2MoS_4$ or $K_3Fe(CN)_6$, DMF/water, 80 ℃, 24h; iv)$H_2$, 480 ℃, 1h, then annealed under argon at 1000 ℃ for 2h. (b) Polarization curves of MoP@PC, FeP@PC, and Pt/C. (c) Tafel plots of MoP@PC, FeP@PC, and Pt/C. Reprinted with permission from ref. 197. Copyright 2015 Wiley-VCH.

Precise control of the metal catalytic site of precursor as well as the location of nitrogen heteroatom is quite important for the performance of resultant carbon materials. The use of metal containing COFs is an option for this purpose. Dai *et al.* reported the use of non-noble metals incorporated COFs as templates and precursors for producing metal/carbon catalyst.[198] The resultant materials with rather uniform metal/nitrogen distribution showed efficient electrocatalytic activities toward 4 electron oxygen reduction reaction in both alkaline and acid media with an excellent stability as well as being free from methanol-crossover/CO-poisoning drawbacks.

## 4 Conclusion and perspectives

NIONs are emerging unique class of porous polymer networks that inherently combine high-density packing of ion pairs with porosity. The abundant synthetic routes and availability of building blocks allow to generate NIONs with rich structural diversity and functions. The rapid growth of this field in the past few years focuses on the design, synthesis and functional exploration of these materials. Especially, when charged character and pores in nanometer range work synergistically, intriguing applications in catalysis, energy storage and conversion, as well as environmental applications are obtained, outperforming the behavior of conventional nanoporous network with neutral skeleton.

As for the design and synthesis of porous structure, the porosity, pore environment and functionality can be tuned. Nevertheless, the reported values of $S_{BET}$ are currently still limited (< 1600 $m^2$ $g^{-1}$) especially in comparison with that of MOFs and carbons. Enhanced surface area is therefore one of the future goals. Besides, delicate control of the pore structure (size and shape) and systematically tuning the synergistic effect of pore size and electrostatic interaction at a molecular scale are not fully understood and deserve more in-depth investigations.

Already now, NIONs by their size or shape selectivity, enhanced mass transport, and special pore environments, enable outstanding activities in heterogeneous catalysis. We believe that the combination of intrinsic ionic characteristic with additional functional groups/sites will offer new opportunities to extend catalytic applications of NIONs, for instance in tandem catalysis and for reaction cascades. It should be mentioned that the most of reported NIONs to date are amorphous structures: the pore structure could not be fully controlled due to the disorder-generating processes of polymerization and crosslinking. Such characteristics lead to a random distribution of catalytic sites in the porous network, which in turn makes the evaluation of catalytic sites and the control of the catalytic performance difficult. Recently, ionic COFs with ordered pore systems might generate new insight in this area, whilst their application in catalysis still needs to be expanded and evaluated. In addition to delicate choice of the raw materials and protocol of polymerization, seeking for appropriate conditions to facilitate weak interactions, such as π－π (cation-π, anion-π) stacking and hydrogen binding, is believed to be helpful to generate the materials with higher/improved pore control. As a final step, one might envision the synthesis of a material with densely packed "artificial enzyme pockets", with all secondary interactions being delicately placed to enable maximum reactivity and/or selectivity.

NIONs have found to be excellent candidates toward environmental related applications in terms of high absorption capacity and kinetics with regard to neutral porous materials. The enhanced performance with fast, efficient capture of pollutants, high selectivity and high recyclability of the absorbents has to be further explored. It is believed that the principle of designed materials with a monoatomic layer thick pore wall, open channels to avoid entrainment and a dense packing of ion exchange sites contributing to a high ion exchange capacity will inspire following research works.[199] In addition, to achieve meaningful adsorption capacity at lower concentrations or high selectivity, a high enthalpy of adsorption ($\Delta H_{ads}$) is required. The demanded enthalpies typically lie well beyond pure physical adsorption processes and will instead involve materials that interact multisite-physically or chemically with the analyte of interest. A few examples have already hinted to the benefits of ionic group triggered coordination/chemically bound guest species, whilst proper control of binding energy (high enough to boost selectivity, but low enough for a possible desorption process) is the key for optimization and recyclability of the absorbents without excessive energy demands. This could turn into a leading descriptor for customization design of NIONs.

NIONs based energy applications is indeed a burgeoning field in materials science, and the already proven practical applications are concerned with their structure stability with regard to mechanical load or chemically harsh conditions *(e.g.*, strong acidic or basic conditions, high electrochemical stress). Although a few NIONs with superior stability have been explored, effective methods to address this issue remain an open race.

A parallel challenge in this area is concerned with the conductivity of framework. Excellent exciton migration and charge carrier transport is necessary for many energy processes. Methodologies and molecular structures that can enhance the charge carrier mobility and electric conductivity are known from the development of OPV systems, but gain here a "third dimension", as packing and neighboring effects become addressable and are highly relevant. In this regard, systematic investigations are demanded to clarify the structure-property relationship of "3D-heterogeneous organic conductors". But even on a more primary level, the development of conjugated crosslinking porous network with built-in high density of ionic pairs is believed to bring new inspiration into materials chemistry of this area. Besides, device integration *via* NIONs designed for good processability is also of great importance. We envision that some novel techniques, such as ink-jet printing, wet-lithography or roll-to-roll processes, might be implemented to achieve these targets of integration and production.

Another emerging trend is downsizing the bulk NIONs to nanodimension, which holds great promise to extend the potential applications. Very recently, an intriguing example was described: guanidinium halide containing ionic network was self-exfoliated into ionic covalent organic nanosheets.[200] Intrinsic charges in the framework backbone and sandwiched anions in between the layers were believed to be responsive for such self-exfoliation process. Compared with conventional on-surface synthesis procedure for neutral COF-derived nanosheets, such method toward spontaneously generated nanosheets highlights the advantage of charge species which enable time-/energy-efficient procedures. The downsized ICONs with reduced dimensionality and well-defined in-plane charge transport may provide unique photoelectrochemical properties.

Currently, the investigation of NIONs is still in its infancy, and only materials of a limited scope have been designed and explored. Considering the versatility of ionic monomers as well as synthetic methods, there is however so much space to produce intriguing structures with improved physico-chemical properties. As a unique type of functional porous organic material with outstanding performances, NIONs will continue to draw interest and enquiry from both academia and industry. It will be exciting to witness the rapid development of this new field in the years to come.

**Acknowledgements**

We would like to thank Max Planck Society for financial support and Prof. Arne Thomas for helpful discussions. J. Y. thanks the ERC (European research council)helfpful Starting Grant (project number 639720-NAPOLI). J. K. S thanks AvH (Alexander von Humboldt) foundation for a postdoctoral fellowship.